\documentclass[11pt,a4paper]{article}

\usepackage{jheppub}
\usepackage{epstopdf}

\usepackage{mathrsfs}
\usepackage{psfrag}
\usepackage{color}
\usepackage{hyperref}

\usepackage{slashed}
\usepackage{simplewick}
\usepackage{cancel}
\usepackage[utf8]{inputenc}

\usepackage[table]{xcolor}

\usepackage{amsmath,bbm,array,amsfonts,graphicx,wrapfig,arydshln,lscape,float,multirow,longtable,rotating,makecell}
\usepackage{url}

\usepackage{caption}
\usepackage{subcaption}

\usepackage[backgroundcolor=black!5, linecolor=green!60]{todonotes}

\hypersetup{pdftitle={},pdfcreator={},linkcolor=[rgb]{0.15,0.35,0.75},colorlinks=true,citecolor=[rgb]{0.675,0,0.2},urlcolor=[rgb]{0.15,0.35,0.65}}
\let\olditemize\itemize\renewcommand{\itemize}{\vspace{-2pt}\olditemize\setlength{\itemsep}{1pt}\setlength{\parskip}{0pt}\setlength{\parsep}{-0pt}}
\let\oldenumerate\enumerate\renewcommand{\enumerate}{\vspace{-4pt}\oldenumerate\setlength{\itemsep}{1pt}\setlength{\parskip}{0pt}\setlength{\parsep}{0pt}}

\renewcommand{\thefootnote}{\fnsymbol{footnote}}
\newpage
\newcommand{\p}[1]{(\ref{#1})}

\newcommand \vev [1] {\langle{#1}\rangle}
\newcommand \ket [1] {|{#1}\rangle}
\newcommand \bra [1] {\langle {#1}|}

\newcommand{\cM}{{\cal M}}

\newcommand{\cN}{{\cal N}}

\newcommand{\cG}{{\cal G}}

\newcommand{\cP}{{\cal P}}
\newcommand{\cQ}{{\cal Q}}

\newcommand{\nt}{\notag\\} 
\newcommand{\pa}{\partial}
\newcommand{\ep}{\epsilon}

\renewcommand{\a}{\alpha}
\renewcommand{\b}{\beta}
\renewcommand{\d}{\delta}
\newcommand{\g}{\gamma}

\newcommand{\la}{\lambda}

\newcommand{\da}{{\dot\alpha}}
\newcommand{\db}{{\dot\beta}}
\newcommand{\dg}{{\dot\gamma}}
\newcommand{\dd}{{\dot\delta}}
\newcommand{\q}{\theta}

\newcommand{\bz}{\bar z}
\newcommand{\tx}{\tilde{x}}
\newcommand{\tr}{\mbox{tr}}

\newcommand\re[1]{(\ref{#1})}

\def\nn{\nonumber}

\def\bom#1{{\mbox{\boldmath $#1$}}}

\csname @addtoreset\endcsname{equation}{section}

\makeatletter
\def\timenow{\@tempcnta\time
  \@tempcntb\@tempcnta
  \divide\@tempcntb60
  \ifnum10>\@tempcntb0\fi\number\@tempcntb
  \multiply\@tempcntb60
  \advance\@tempcnta-\@tempcntb
  :\ifnum10>\@tempcnta0\fi\number\@tempcnta}
\makeatother

\allowdisplaybreaks[3]

\makeatletter
\renewcommand\@fpheader{} 
\renewcommand\@journal{}
\makeatother

\title{
From correlation functions to event shapes in QCD
}

\preprint{LAPTH-003/20, MPP-2020-8}

\author{D.\ Chicherin$^a$, J.\ M.\ Henn$^{a}$, E.\ Sokatchev$^{b}$, K.\ Yan$^{a}$}

\affiliation{
$^a$ Max-Planck-Institut f{\"u}r Physik, Werner-Heisenberg-Institut, 80805 M{\"u}nchen, Germany\\
$^b$ LAPTh, Universit\'e Savoie Mont Blanc, CNRS, B.P. 110, F-74941 Annecy-le-Vieux, France}

\emailAdd{chicheri@mpp.mpg.de}
\emailAdd{henn@mpp.mpg.de}
\emailAdd{emeri.sokatchev@cern.ch}
\emailAdd{kyan@mpp.mpg.de}

\abstract{
We present a  method for calculating event shapes in QCD based on correlation functions of conserved currents. 
The method has been previously applied to the maximally supersymmetric Yang-Mills theory, but we demonstrate that supersymmetry is not essential. As a proof of concept, we consider the simplest example of a charge-charge correlation at one loop (leading order). We compute the correlation function of four electromagnetic currents and explain in  detail the steps needed to extract the event shape from it. The result is compared to the standard amplitude calculation. The explicit four-point correlation function may also be of interest for the CFT community.}

\begin{document}

\maketitle

\thispagestyle{empty}

\bigskip
\bigskip

\setcounter{page}{1}\setcounter{footnote}{0}
\pagestyle{plain}
\renewcommand{\thefootnote}{\arabic{footnote}}

\setcounter{page}{1}\setcounter{footnote}{0}

\section{Introduction}

Event shapes are important infrared safe observables in QCD  \cite{Fox:1978vu,Basham:1978bw,Basham:1978zq,Ellis:1980wv,Kunszt:1989km,Kunszt:1992tn,Biebel:2001dm,Belitsky:2001ij}.
For example, energy weighted cross sections can be used to measure the
strong coupling constant, or to study jet physics. As such, they have the potential
of connecting partonic and hadronic cross sections, especially within approaches 
that are valid non-perturbatively. One of the best known event shapes
is the energy-energy correlation (EEC), widely studied in, e.g., $e^+ e^-$ annihilation.

There has been considerable recent theoretical interest in event shapes.
While numerical results are in principle enough for comparison with experiment,
they may require heavy computer resources, or have intrinsic limitations
as far as numerical accuracy is involved. Analytic results, on the other hand, are typically
fast to evaluate to essentially arbitrary numerical precision.

To illustrate some of the recent analytic developments, let us mention that
the next-to-leading-order  (NLO) result for the EEC in QCD was obtained analytically in \cite{Dixon:2018qgp}, 40 years after the leading-order (LO) result \cite{Basham:1978bw,Basham:1978zq} (the numerical result had been known earlier \cite{Kunszt:1992tn,Glover:1994vz}). Remarkable progress was achieved in maximally supersymmetric Yang-Mills ($\cN=4$ sYM) theory, where the first ever NLO result was obtained in \cite{Belitsky:2013ofa}. Moreover,  while the EEC at NNLO is known only numerically  in QCD \cite{DelDuca:2016ily}, even an analytic NNLO result is available in the  $\cN=4$ sYM theory \cite{Henn:2019gkr}. 
This progress was made possible by a novel approach  \cite{Hofman:2008ar,Belitsky:2013xxa,Belitsky:2013bja} that will also be central to this paper.

Furthermore, analytic approaches are much better suited to studying observables in
extreme kinematic limits, such as the back-to-back limit or small angle limit,
where large logarithms occur that need to be resummed 
in order to compare to experiment. Much recent progress has also been made in this direction, see \cite{Korchemsky:2019nzm,Dixon:2019uzg,Kologlu:2019mfz}.

The traditional approach to computing event shapes uses amplitude methods and assembles the 
weighted cross sections from various ingredients, such as real and virtual contributions
and phase-space integrals.   
While the final
result is infrared finite, the intermediate expressions are infrared divergent and require intricate cancellations of singularities. The phase space integrals involved are often difficult to evaluate. 
In contrast, the novel approach 
is based on finite correlation functions. For example, in the case of the EEC, one starts with  
the correlation function of two currents representing the sources, and two energy-momentum
tensors representing the detectors. The event shape is then obtained by sending the detectors to lightlike infinity and integrating over their working time.
This is a conceptual improvement, as unnecessarily large intermediate expressions (involving unphysical regulated terms) are avoided.

Furthermore, this approach has the potential of exposing new structural
properties of the observables: thanks to the connection between the two objects, the general properties
of the four-point functions can inform us about the behavior of the event shapes, in particular beyond perturbation theory.
For example, it was shown in \cite{Kologlu:2019mfz}
that the light-ray OPE gives a non-perturbative expansion for event shapes 
in terms of  conformal blocks, and starting from OPE data they 
were able to make new four-loop predictions for the small angle expansion.
Recently, a new integral representation
for the EEC in the  $\cN=4$ sYM theory was found in  \cite{Henn:2019gkr}. It relates the EEC to a two-fold integral of the triple discontinuity
of the four-point correlation function. 
In this way,  information on the analytic properties of the correlation functions can be used  
to derive consequences for the event shapes.

In the literature, the correlation function approach was first proposed  for energy and charge correlations in a generic conformal field theory in  \cite{Hofman:2008ar}. It was  developed further in \cite{Belitsky:2013xxa} and applied to these two event shapes  (as well as to the mixed energy-charge correlation)  
in the particular framework of the $\cN=4$ sYM theory in \cite{Belitsky:2013bja,Belitsky:2013ofa}. 
One should bear in mind that  this theory is very special due to the extended super(conformal) symmetry, so 
one may doubt how useful the approach is in realistic QCD calculations. Moreover, supersymmetry
relates different event shape observables, leaving  no essential difference (a prefactor) between
energy-energy and charge-charge correlations \cite{Belitsky:2013bja,Belitsky:2014zha,Korchemsky:2015ssa}.

In this paper, we overcome these limitations. We work directly in QCD, and compute for the first
time an event shape at one loop (or LO, i.e. order $O(g^2)$ in the coupling) using the correlation function approach.
Compared to other calculations, we observe that while supersymmetry has been helpful in the past and provided simplifications, 
it is not necessary for the method to work. Our main goal is the proof of a new concept, therefore we chose the simplest example of a charge-charge correlation (QQC).  The method applies equally well to the EEC, although the computation of a correlation function with energy-momentum tensors is technically more involved. 

This paper constitutes a first application of the new method to QCD, and therefore we also performed
the calculation using standard amplitude methods, as a check. 
It should be emphasized that at the order we work in, the standard amplitude
calculation is certainly more efficient. As we discuss in the outlook section, there are reasons to think that
this will change at higher orders.

As a key ingredient of our analysis, we obtain the four-point correlation function 
of spin-one operators (electromagnetic  currents) at one loop in QCD. This object is interesting in itself because the conserved currents are protected from UV renormalization. 
At the order $O(g^2)$ the correlation function is conformally covariant because the beta function is of order $O(g^3)$. Conformal four-point functions of operators of spin zero are well studied.
They are very important in the context of the operator product expansion (OPE),
and in particular for the conformal bootstrap.
The general structure of conformal correlators of operators with spin has been discussed
in, e.g., refs.  \cite{Sotkov:1980qh,Dymarsky:2013wla,Kravchuk:2016qvl,DK}.
As far as we are aware, our result is the first explicit loop-level  non-supersymmetric example
of such a four-point correlator.
In principle, it could also be determined by the CFT data, i.e.
the scaling dimensions and the structure constants, but this is yet to be 
worked out. We believe that our correlation function  can serve as a first data point for OPE studies.

The paper is organized as follows. In  Section \ref{s1} we review the definition of event shapes 
as weighted cross sections, and how they can be obtained from correlation functions.
The following Section \ref{sec2} is dedicated to the calculation of the correlation function of four currents at one loop in QCD using  the Lagrangian insertion technique. The result for the four-point correlator, which may be of interest in itself
to researchers in CFT, is given in eq.~(\ref{JJJJ2}).
In Section \ref{s4} we extract from this result  the charge-charge correlation at LO in QCD, eq.~\p{final}, 
and we compare with the result of the traditional amplitude calculation, which can be found in Appendix \ref{ampcalc}.
Finally, we conclude  and comment on future directions in Section \ref{s5}.

\section{Event shapes from correlation functions}\label{s1}

In this section we review the basic definition of an event shape in  QCD and its relationship to correlation functions. We restrict ourselves to the case of the electromagnetic current $J_\mu$ as the source and the associated electric charge $\cQ$ as the detector. The generalization to other setups is straightforward \cite{Hofman:2008ar,Belitsky:2013bja}. 

\subsection{Event shapes as weighted cross sections}\label{s1.1}

 Let $P\cdot J(x)= P^\mu J_\mu(x)$ be a vector current $J_\mu(x)= \tr\bar\Psi\g_\mu\Psi$ projected with a (complex) polarization vector $P^\mu$. Here $\Psi$ denotes a set of Dirac spinors describing the quarks and antiquarks in the fundamental representation of the color group. Further, let $\ket{X}$ be the final state created by this gauge invariant operator   from the vacuum.   To lowest order in the coupling it consists of  
a quark-antiquark  pair. In general, the state $\int d^4 x e^{iqx} P\cdot J(x)\ket{0}$ involves an arbitrary number of particles, with total momentum $q^\mu$ and zero color charge.
The amplitude for the creation of a particular final state $\ket{X}$ with total momentum $k_X$  is
\begin{align}
 \vev{X| \int d^4 x e^{iqx} P\cdot J(x)|0} =  (2\pi)^4  \delta^{(4)}(q-k_X)\mathcal M_{J\to X}\,,
\end{align}
where
\begin{align}\label{M}
 \mathcal M_{J\to X}=\vev{X|  P\cdot J(0)|0} \,
\end{align}
 is the form factor of the current on the on-shell state $X$. 
The total probability
of this process is given by the sum over all the final states,  
\begin{align}\label{tot}
\sigma_{\rm tot}(q) = \sum_X (2\pi)^4 \delta^{(4)}(q-k_X)|\mathcal M_{J\to X}|^2\,.
\end{align}
Inserting the completeness relation
\begin{align}\label{corel}
\sum_X \ket{X}\bra{X}=1\,,
\end{align}  
we can rewrite \re{tot} as
\begin{align} 
\sigma_{\rm tot}(q)  & = \int d^4 x\, e^{iqx}   \vev{0|  \bar P\cdot  J(x)\   P\cdot J(0)|0}_W \,. \label{2.8}
\end{align}
In other words, the total cross section can be interpreted as the Fourier transform of the two-point correlation function of the current, $\vev{J_\mu(x) J_\nu(0)}_W$, projected with the polarization matrix $\bar P^\mu P^\nu$. It is important to point out that this correlation function, defined in Minkowski space-time, is not time-ordered. This is denoted by the subscript $W$ meaning Wightman prescription. The alternative definition \p{2.8} of the total cross section \p{tot} allows us to avoid infrared divergences at the intermediate steps and the necessity to
sum over all the final states.

An event shape is defined as a weighted cross section  with a weight factor $w(X)$ for the contribution of each state $\ket{X}$,
\begin{align}\notag\label{wcs}
\sigma_{w}(q) &= \sigma_{\rm tot}^{-1}\sum_X (2\pi)^4 \delta^{(4)}(q-k_X)w(X)|\mathcal M_{J\to X}|^2 
\\
 & = \sigma_{\rm tot} ^{-1} \int d^4 x\, e^{iqx}  \sum_X   \vev{0|  \bar P\cdot  J(x) |X}  w(X)
\vev{X|  P\cdot J(0)|0} \,,
\end{align}
 normalized so that $\sigma_{w}(q)=1$ for
$w(X)=1$.  The event shape $\sigma_{W}(q)$ describes the flow of the
quantum numbers of the particles, i.e. energy,  charges, etc.  In this paper we study the case of charge flows. 

For a given final state $\ket{X} = \ket{k_1,\dots,k_\ell}$, consisting of $\ell$ massless particles, $k_i^2=0$, with charges $Q_i$ and 
total momentum $\sum_i k_i^\mu = q^\mu$, the weight factor is defined in the rest frame of the source $q^\mu=(q^0,\vec 0)$ as 
\begin{align}\label{w-energy}
w_{\mathcal Q}(k_1,\dots,k_\ell) = \sum_{i=1}^\ell Q_i \,\delta^{(2)}(\Omega_{\vec k_i} - \Omega_{\vec n}) \,,
\end{align}
where   $k_i^\mu=(k_i^0,\vec k_i)$  and $\Omega_{\vec k_i}=\vec k_i/|\vec k_i|$ is the solid angle in the direction of $\vec k_i$. The unit vector $\vec n$ (with $\vec n^2=1$)   indicates the  direction of the charge flow.
The weight  \re{w-energy} is the eigenvalue of the charge flow operator,
\begin{align}\label{E-spectr}
\mathcal Q(\vec n) \ket{X}=w_{\mathcal Q}(X)   \ket{X}\,.
\end{align}
The explicit expression for the operator $\mathcal Q(\vec n)$ is given in terms of
the time component of the current  as an integral over the working time $t$ of the detector at the point  $r\vec n$ infinitely far away from  the collision point \cite{Ore:1979ry,Sveshnikov:1995vi,Korchemsky:1997sy,Korchemsky:1999kt,BelKorSte01} (see also \cite{Hofman:2008ar}) :
\begin{align} \label{Q-flow}
\mathcal Q(\vec n) =  \int_0^\infty dt \, \lim_{r\to\infty} r^2\, n^i\, J_{0 }(t,r\vec n)\,.
\end{align}
 It satisfies the commutativity condition
\begin{align}\label{Q-com}
 [  \mathcal Q(\vec n) ,  \mathcal Q(\vec n') ] = 0 \qquad  {\rm for} \ \vec n \neq \vec n'\,, 
\end{align}
meaning that the charge flows in the directions $\vec n$ and $\vec n'$  can be measured separately. 

Making use of \re{Q-com} we can define a weight 
which measures the charge flows in two (or more) directions
$\vec n, \vec n'$ simultaneously: 
\begin{align}\label{QQ}
 \mathcal Q(\vec n)  \mathcal Q(\vec n')\ket{X}=w_{\mathcal Q(\vec n)}(X)    w_{\mathcal Q(\vec n')}(X) \ket{X}
 \,.
\end{align}
Substituting eqs.~\p{E-spectr} and \p{QQ}  into \re{wcs} we can apply the completeness relation \p{corel} and obtain the following representation of 
the corresponding weighted cross sections (or charge flow correlations)
\begin{align}\label{sigma-Q}
&\vev{\mathcal Q(\vec n)}   
= \sigma_{\rm tot}^{-1} \int d^4 x\, e^{iqx}  \vev{0|  \bar P\cdot  J(x) \, \mathcal Q(\vec n) \, P\cdot J(0)|0}_W\,, \\
&\vev{\mathcal Q(\vec n)  \mathcal Q(\vec n')}     
= \sigma_{\rm tot}^{-1} \int d^4 x\, e^{iqx}  \vev{0|  \bar P\cdot  J(x) \, \mathcal Q(\vec n)  \mathcal Q(\vec n')\, P\cdot J(0)|0}_W \,, \quad  \text{etc.} \label{sigma-QQ}
\end{align}
We repeat that the product of operators in \p{sigma-Q}, \p{sigma-QQ}  is not time ordered and their correlation functions are of the Wightman type.  We will refer to the currents at points $x$ and $0$ as to the sink and source, respectively. The flow operators $\cQ$ will be called detectors. The event shapes \p{sigma-Q}, \p{sigma-QQ}, etc. are called single charge, charge-charge, etc. correlations. 

In this paper we specialize to the lowest nontrivial perturbative level (LO), i.e. $O(g^2)$.\footnote{At Born level $O(g^0)$ the event shape is given by contact terms, which we do not consider in  this paper.  } The standard calculation of the  weighted cross section \p{sigma-QQ} from amplitudes  is given in Appendix~\ref{ampcalc}. The bulk of the paper is devoted to obtaining the same result from the integrated correlation function of four electromagnetic currents. 

\subsection{Weighted cross sections from correlation functions}  \label{s1.2}

The underlying quantity in the definitions \re{sigma-Q}, \p{sigma-QQ}, etc.  of the weighted cross sections of $m$ charge flow detectors $\cQ(\vec n_1), \ldots, \cQ(\vec n_m)$  are the Wightman correlation functions of $(m+2)$ currents  $\vev{J_\mu(x) J_{\la_1}(x_1) \ldots J_{\la_m}(x_m) J_\nu(0)}_W$. In this subsection we explain how the Wightman correlation function in  \re{sigma-Q}  
can be obtained from its Euclidean counterpart by analytic continuation. We then apply it to computing the single-charge correlation at Born level (free theory). This simple example illustrates the procedure that we implement in the rest of the paper. At the end of the subsection we comment on the non-trivial analytic continuation of the four-point correlation function at loop level. 

The Euclidean correlation functions have short-distance singularities when  $x_i\to x_j$. In Minkowski space, additional singularities arise when the
operators become lightlike separated, $x^2_{ij}\to0$. In this case the analytic properties of the correlation function depend on the 
ordering of the operators.  

In \re{Q-flow} we  defined the charge flow operator in the rest frame of the source
$q^\mu=(q^0,\vec 0)$. The detector 
coordinates  $x^\mu=(t,r\vec n)$ can be decomposed in the basis of  two lightlike vectors,
\begin{align}\label{x}
& x^\mu = x_+ n^\mu + x_- \bar n^\mu \,,\qquad n^\mu=(1,\vec n)\,,\qquad \bar n^\mu=(1,-\vec n)\,.
\end{align} 
Manifest Lorentz covariance is restored by independent rescaling of these vectors,
\begin{align}\label{rhon}
n^\mu \to \rho\, n^\mu\,, \qquad \bar n^\mu \to \bar\rho\, \bar n^\mu\,, \qquad \rho, \bar\rho >0 \,. 
\end{align}
Then  the covariant definition of the light-cone coordinates in \p{x} becomes
\begin{align}\label{xpm}
& x_+={(x\bar n)\over (n \bar n)}\,,\qquad x_-  ={(xn)\over (n \bar n)} \, .
\end{align}

We can now reformulate the detector limit $r\to\infty$ and the integration over the working time interval  $0\le t <\infty$ in terms of the light-cone variables $x_{\pm}$ (for the detailed physical motivation see \cite{Hofman:2008ar,Belitsky:2013xxa,Belitsky:2013bja}),
\begin{align}\label{E-new}
\mathcal Q(n) =  \int_{-\infty}^\infty  dx_- (n\bar n)   \lim_{x_+\to\infty} x_+^2 \, J_{+}(x_+ n  + x_- \bar n)\,.
\end{align}
Here $J_+\equiv \bar n^\mu  J_\mu(x)/(n\bar n)$ is the covariant light-cone projection of the current.  Lorentz covariance requires that the charge flow operator transforms homogeneously under the rescaling \p{rhon}, $\cQ \to \rho^{-2} \cQ$. As we shall see later on, the dependence on the auxiliary vector $\bar n$ is redundant and it drops out of the final result for the event shape.

\subsection{A simple example: single charge correlation}\label{s1.3}

Here we present a very simple example which illustrates all the main steps in obtaining an event shape from a Euclidean correlation function of currents. Consider the three-point function of currents made from a single Dirac fermion in the free theory (Born) approximation. According to \p{sigma-Q}, the single charge correlation is given by\footnote{ In order to have a non-vanishing correlation function we need to consider different currents. Here we chose a  vector current $J^{(v)}$ at points $x_1$ and $0$ as the sink/source, and an axial current $J^{(a)}$ at point 2 as the detector. Another possibility is to have three vector currents with antisymmetrized flavor indices. } 
\begin{align}\label{det-lim}
\vev{\mathcal Q(\vec n)} &= \sigma_{\rm tot}^{-1} \int d^4 x\, e^{iqx_1}  \int_{-\infty}^\infty  dx_{2-} (n\bar n)\nt
& \times    \lim_{x_{2+}\to\infty} x_{2+}^2 \,  \vev{0|  \bar P\cdot J^{(v)}(x_1) \, \bar n\cdot J^{(a)}(x_{2+} n  + x_{2-} \bar n) \, P\cdot J^{(v)}(0)|0}_W\,.
\end{align}

Using  the  Feynman rules from Section~\ref{sec2} we can easily compute the Euclidean correlation function. We find  (up to a normalization factor)
\begin{align}
G_E  &=\vev{ \bar P\cdot J^{(v)}(x_1)\ \bar n\cdot J^{(a)}(x_2)\ P \cdot J^{(v)}(0)}_E\nt
& \sim  \frac{1}{(x_{12}^2 x_{1}^2 x_{2}^2)^2} \Big[\tr(\bar P \tx_{12} \bar n  \tx_2 P \tx_1) + \tr(\bar P \tx_{1} P \tx_2\bar n \tx_{21})\Big]\,.
\label{JJJtree}
\end{align}
We need to perform the analytic continuation to the Wightman   function in Minkowski space. To this end we replace each Euclidean interval by a Minkowski one, with the prescription
\begin{align}\label{1.19}
x^2_{ij} \ \rightarrow \ x^2_{ij} -i\ep x^0_{ij} \quad \text{for} \ i<j\,. 
\end{align}
The next step is to take the detector limit of the Wightman function. The detector coordinate is $x_2^\mu = x_{2+} n^\mu + x_{2-}\bar n ^\mu$. Then, for $x_{2+}\to\infty$
we have
\begin{align}\label{1.20}
&x^2_{12} -i \ep  x_{12}^0\ \to\ 2x_{2+} (  x_{2-} (n\bar n)-(x_{1}n) +i \ep)\,, \nt
& x^2_{2} -i\ep x_{2}^0\ \to\ 2x_{2+} (x_{2-} (n\bar n)- i\ep)   \,,
\end{align} 
and we find from \re{JJJtree}  
\begin{align}\label{4.11}
\lim_{x_{2+} \to \infty}  x^2_{2+} G_W   \sim \frac{(n\bar n)\, i \ep_{\mu\nu\la\rho} \bar P^\mu  P^\nu n^\la x_1^\rho}{(x_{2-}(n\bar n)-(x_{1}n)+i\ep)^2 (x_{2-}(n\bar n) - i\ep)^2(x^2_1 -i\ep x^0_1)^2}\,.
\end{align}

According to \re{det-lim}, we have to integrate this expression over the detector time (or  light-cone coordinate) $x_{2-}$.   In \p{4.11} we see  two double poles  in $x_{2-}$ located on the two sides of the real axis. Closing the integration contour 
in the upper half-plane, we get
\begin{align}\label{det3}
 \int_{-\infty}^\infty dx_{2-} (n\bar n)  \lim_{x_{2+} \to \infty}  x^2_{2+} G_W   \sim \frac{ i \ep_{\mu\nu\la\rho} \bar P^\mu  P^\nu n^\la x_1^\rho}{ ((x_1n) -i\ep)^3 (x_1^2 -i\ep x_1^0)^2}\ .
\end{align}
As expected, the auxiliary lightlike vector $\bar n$ has dropped out, and the result is homogeneous of degree $(-2)$ under the rescaling of the vector $n$. The Levi-Civita tensor  originates from  the parity-odd three-point correlation function that we started with. 

The last step is the Fourier integral in \p{det-lim}, computed with the help of the formula 
\begin{align}\label{theta} 
&\int \frac{d^{4} x \,  e^{i q  x}}{(  x^2 - i \ep x^0 )^{a+1}   ( (nx) - i \ep )^{b+1}  }  =   \frac{ (4 \pi)^{3} \, ( i/2)^{2a +b + 3 } }{ \Gamma(a+1) \Gamma(a +b +1 )  }  \theta(q^2) \theta(q^0)     \frac{(q^2)^{a+b } }{ (nq )^{b+1} }  \,.
\end{align}
Due to the current conservation condition we can identify $P_\mu \to P_\mu + \la q_\mu$. This allows us to choose a  polarization vector orthogonal to the momentum, $(Pq)=(\bar Pq)=0$. 
The total cross section in \p{det-lim} is the Fourier transform of the two-point function of the source,
\begin{align}\label{}
\int d^4x\, \vev{\bar P\cdot J P\cdot J} \sim  \theta(q^2) \theta(q^0)   (\bar PP) q^2 \,.
\end{align}
 In the rest frame  $q^\mu=(q^0,\vec 0)$ we have $P^\mu=(0,\vec p)$ and $(\bar PP)= - |\vec p|^2$. Finally, the event shape is given by the expression
\begin{align}\label{}
\vev{\mathcal Q(\vec n)} \sim \frac{i}{|\vec p|^2} \, \ep_{ijk} p^*_i p_j  n_k\,,
\end{align}
which  exists only for a complex polarization vector \cite{Hofman:2008ar}.

This concludes our pedagogical example. In Sect.~\ref{s4} we apply the same procedure to the four-point correlation function of electromagnetic currents in QCD. 

\subsection{Charge-charge correlation for  $e^+ e^-$ annihilation}\label{s1.4}

\begin{figure}
\begin{center}
\includegraphics[width=5.5cm]{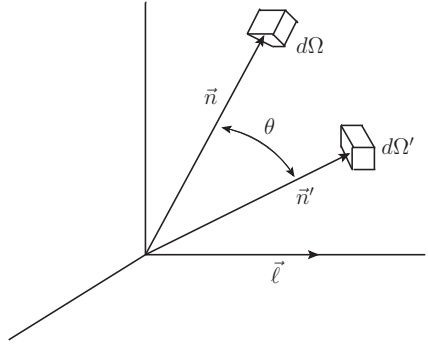}
\end{center}
\caption{Two-detector configuration in $e^+e^-$ scattering. The vectors $\vec n, \vec n', \vec\ell$ denote the directions of the detectors and of the $e^+ e^-$ beam, respectively.}
\label{2det}
\end{figure}

Let us recall the physics of the collider experiment $e^+ e^- \to \gamma^*(q) \to X$. A pair of leptons $e^+, e^-$ annihilates into a virtual photon $\g*$ with off-shell momentum $q^\mu$, which in turn decays into a number of quarks and gluons  (final state $X$). The matrix element squared for this process can be written in the factorized form
\begin{align}\label{e225}
\sum_X |\cM_{e^+ e^- \to X}|^2 = L_{\mu\nu} H^{\mu\nu}\,. 
\end{align}
The (symmetric) hadronic matrix $H$ accounts for the non-trivial decays $\gamma^* \to X$. 
In the case of  charge-charge correlations under consideration, the charge-weighted  hadronic matrix is identified with the  correlation function of two currents and two charge operators \p{E-new},
\begin{align}\label{}
H_{\mu\nu}(q,n,n') = \int d^4 x\, e^{iqx}  \vev{0|  J_\mu(x) \, \mathcal Q( n)  \mathcal Q( n')\,  J_\nu(0)|0}_W\,. 
\end{align}
The leptonic matrix $L$ is made from two vertices $e^+ e^- \to \gamma^*$. In the typical case of an unpolarized beam (i.e. summing over the polarizations of the incoming particles) it has the following simple form \cite{Ellis:1991qj}:
\begin{align}\label{e227}
 L_{\mu\nu} \sim \ell^+_\mu \ell^-_\nu +  \ell^+_\nu \ell^-_\mu  -  \ell^+\cdot \ell^-  \eta_{\mu\nu} \,.
\end{align}
Here  $\ell^\pm$ (with $(\ell^\pm)^2=0$ and  $\ell^+ + \ell^-= q$) are the on-shell momenta of the two leptons. The current conservation conditions $q^\mu L_{\mu\nu}= q^\mu H_{\mu\nu}= 0$  become $L_{0\nu}=H_{0\nu}=0$ in the center-of-mass frame $q^\mu=(q^0,\vec 0)$. In it $\ell^\pm =(q^0/2, \pm \vec \ell)$ where the vector $\vec\ell$ defines the direction of the beam. Then \p{e225}, \p{e227} yield
\begin{align}\label{e228}
\sum_X |\cM_{e^+ e^- \to X}|^2  \sim \left[ \frac{(q^0)^2}{4} \delta_{ij} + \vec \ell^2  \delta_{ij} +2 \ell_i \ell_j \right] H^{ij}(q^0, \vec n, \vec n') \,. 
\end{align} 
The  term in the brackets contains the information about the beam direction, relative to the directions of two detectors $\vec n, \vec n'$ (see Fig.~\ref{2det}). 
 In some of the early papers  \cite{Basham:1978bw,Basham:1978zq} the energy-energy correlation (EEC) was defined as a function of these three directions. Nowadays one considers a simplified  observable (see, e.g., \cite{Fox:1978vu,Ellis:1980wv}), in which one integrates over the direction of the beam $\vec\ell$, or equivalently, over the orientation of the detector plane relative to the beam. Rotation symmetry then tells us that the right-hand side of \p{e228} is reduced to the trace $H^{ii}$. The Lorentz covariant version of this averaging procedure is 
\begin{align}\label{}
\int d^2\Omega_{\ell}\, \sum_X |\cM_{e^+ e^- \to X}|^2  \sim \eta_{\mu\nu} H^{\mu\nu} =   \int d^4 x\, e^{iqx}  \vev{0|  J^\mu(x) \, \mathcal Q( n)  \mathcal Q( n')\,  J_\mu(0)|0}_W\,. 
\end{align}

In conclusion, the charge-charge correlation \p{sigma-QQ}, averaged over the orientation of the detector plane,  becomes
\begin{align}\label{124}
&\vev{\mathcal Q( n)  \mathcal Q( n')}  
= \sigma_{\rm tot}^{-1} \int d^4 x\, e^{iqx}  \vev{0|  J^\mu(x) \, \mathcal Q( n)  \mathcal Q( n')\,  J_\mu(0)|0}_W \,.
\end{align}
 This quantity depends on the virtual photon momentum $q$ and on the lightlike directions $n,n'$ of the two detectors.  Counting the dimensions of the operators involved in \p{124} and taking into account the scaling properties under \p{rhon}, we conclude that
\begin{align}\label{zetadef}
\vev{\mathcal Q( n)  \mathcal Q( n')} = \frac{F_{QQC}(\zeta)}{4\pi^2 (nn')^2}\, \,, \qquad \zeta = \frac{q^2(nn')}{2(qn)(qn')}\,. 
\end{align}
In the center-of-mass frame  the variable  $\zeta$ is related to the angle between the two detectors,
\begin{align}\label{1.27}
2\zeta=(nn')=1-\vec n \cdot \vec n' \equiv 1-\cos\q \,.
\end{align} 
The purpose of the rest of the paper is to compute the event shape function $F_{QQC}(\zeta)$ at the leading order $O(g^2)$ in massless QCD. 

Before we move on, a comment is due on the analytic continuation of the four-point function in \p{124}. At loop level the Euclidean correlator involves a non-trivial function with branch cuts (see \p{Phi1} below). Its analytic continuation may seem to be a highly nontrivial task. A solution was found in \cite{Belitsky:2013xxa,Belitsky:2013bja} (following an idea of G. Mack \cite{Mack:2009mi}). It makes use of the Mellin representation of the function \p{Phi1}. Under the sign of the Mellin integral the analytic continuation is straightforward, according to the rule \p{1.19}. We implement this procedure in Sect.~\ref{se321}. An alternative, even more efficient way  is to replace the detector time integrations by the double discontinuity of the function \p{Phi1} \cite{Caron-Huot:2017vep,Alday:2017vkk,Henn:2019gkr}. This method is applied in Sect.~\ref{se322}.

\section{Correlation function of four vector currents at one loop}\label{sec2}

In this section we calculate the four-point correlation function of electromagnetic currents in a massless gauge theory in the one-loop approximation. At this perturbative level non-abelian effects play no role, so our calculation is valid in QED as well as in QCD. The difference is only in the overall color factor. 

The QCD Lagrangian contains gauge bosons (gluons) and fermions (quarks):\footnote{Gauge theories involving scalar fields, such as $\cN=4$ sYM, are beyond our considerations.}  
\begin{align} 
L =  - \frac{1}{2} \tr \, F_{\mu \nu} F^{\mu \nu} + \bar\Psi i \gamma^\mu {\cal D}_{\mu} \Psi\,. \label{Lagrangian}
\end{align}
Here ${\cal D}_\mu = \pa_\mu - i g A_\mu$ is the covariant derivative with the gauge connection $A_{\mu} = A^a_{\mu} T_a$; $F_{\mu \nu} = \frac{i}{g}[{\cal D}_\mu , {\cal D}_\nu]$ is the field-strength tensor; $\Psi$ is a Dirac spinor  in the fundamental representation of the color group; the color generators of $SU(N_c)$ are normalized as $\tr(T_a T_b) = \frac{1}{2}\delta_{ab}$. 

The   vector $J^{(v)}_{\mu} =\bar\Psi\gamma_\mu \Psi$
and axial vector $J^{(a)}_{\mu} = \bar\Psi\gamma_\mu\gamma_5 \Psi$ currents are classically conserved and are associated with Noether charges. The conservation of the vector current $J^{(v)}_{\mu}$ takes place in the quantum theory as well. The corresponding composite operator is protected from  infinite UV renormalization in perturbation theory. 
The conservation of the axial current at the quantum level is spoiled by the Adler-Bardeen anomaly.

The perturbative calculations in the gauge theory \p{Lagrangian} usually require UV renormalization of the fields and coupling constant. However, in the one-loop approximation $O(g^2)$  we can avoid these complications. Indeed, the interaction vertex renormalization effects play no role since $\beta(g) = {O}(g^3)$. Also,  in our  scheme the fermion propagator renormalization is finite at the one-loop level (see \p{2.22} below). Thus the correlation function of currents in the one-loop approximation\footnote{Here and in the following the terms `tree' and `one-loop' do not refer to the topology of the Feynman graphs representing the correlator. They follow the analogy with the amplitude calculations and denote the perturbative corrections at  orders $O(g^0)$ and $O(g^2)$, respectively.} 
\begin{align}
\vev{J_\mu(x_1) \ldots J_\nu(x_n)}= 
G^{\text{tree}}_{\mu\ldots\nu}(x_1,\ldots,x_n) + g^2 \, G^{\text{1-loop}}_{\mu\ldots\nu}(x_1,\ldots,x_n) + O(g^4) \label{pert-exp}
\end{align}
is a finite four-dimensional quantity which does not require UV regularization. This is crucial for maintaining the conformal invariance of the classical Lagrangian at this perturbative level. As we show below, conformal invariance greatly facilitates our task. 

In the following we use the two-component Lorentz spinor index notation, see Appendix~\ref{conv}. We split the Dirac fermion in a pair of  Weyl (or equivalently Majorana) fermions $\chi$ and $\psi$,
\begin{align}
\Psi = (\chi_{\a} \, ,\, \bar\psi^{\da}) \;\;\;,\;\;\;\;\;\;
\bar\Psi = (\psi^{\a} \, , \, \bar\chi_{\da}) \,.
\end{align}
Then the Lagrangian \p{Lagrangian} in the two-component notation takes the following form 
\begin{align}
L = L_{\rm YM} + i \psi^{\a} {\cal D}^{+}_{\a\da} \bar\psi^{\da} + i \chi^{\a} {\cal D}_{\a\da}^{-} \bar\chi^{\da} \,, \quad 
L_{\rm YM} \equiv  - \frac{1}{2} \tr \, F_{\mu \nu} F^{\mu \nu}\,,  \label{Lexpl}
\end{align}
where ${\cal D}^{\pm}_{\a\da} = \sigma^\mu_{\a\da} {\cal D}^{\pm}_{\mu} = \pa_{\a\da} \mp i g A_{\a\da} $, so that the Weyl fermions $\psi$ and $\chi$ have opposite charges. The vector and axial vector currents
\begin{align}
& J^{(v)}_{\mu} = \psi^{\a} \sigma^\mu_{\a\da} \bar\psi^{\da} - \chi^{\a} \sigma^\mu_{\a\da} \bar\chi^{\da} \,, \qquad J^{(a)}_{\mu} = \psi^{\a} \sigma^\mu_{\a\da} \bar\psi^{\da} + \chi^{\a} \sigma^\mu_{\a\da} \bar\chi^{\da} \label{currentsdef}
\end{align}
are independent linear combinations of the two Weyl currents.  In the following we consider correlation functions of the currents $J_{\a\da} \equiv \frac{1}{2}\sigma^{\mu}_{\a\da} J_\mu = \psi_{\a} \bar\psi_{\da}$ built out of one of the Weyl fermions.
Since neither the interaction vertices nor the propagators mix the two kinds of fermions,  
we can immediately infer the correlation functions involving $J^{(v)}$ and/or $J^{(a)}$.  

\subsection{Computation of the one-loop correlation function via the  Lagrangian insertion procedure}

\begin{figure}
\begin{center}
\includegraphics[width=5cm]{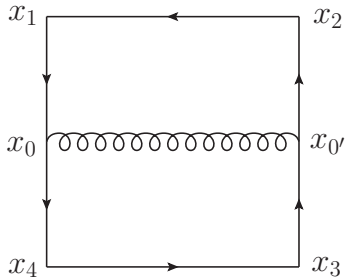}
\end{center}
\caption{A typical Feynman graph of the correlator $\vev{JJJJ}$ at order $O(g^2)$. It contains two interaction vertices corresponding to integrations $\int d^4 x_0 d^4 x_{0'}$.}
\label{difficult}
\end{figure}

The Feynman diagram calculations of correlation functions  in coordinate space   are quite different from the  amplitude calculations in momentum space. In coordinate space we integrate over the space-time position of the interaction vertices. The typical Feynman graph depicted in Fig.~\ref{difficult} contributes to the four-point correlator of currents at order $O(g^2)$. The integrations over $x_0$ and $x_{0'}$ make this graph  comparable to a two-loop amplitude Feynman graph. Fortunately, we can considerably simplify the task via the Lagrangian insertion method \cite{Eden:2000mv,Eden:2010zz}.

We start by rescaling the gauge field,
$A_\mu \to \frac{1}{g} A_\mu$, after which the Lagrangian \p{Lagrangian} takes the following form
\begin{align}
L = \frac{1}{g^2}L_{\rm YM}  + i \psi^{\a} {\cal D}^+_{\a\da} \bar\psi^{\da} + i \chi^{\a} {\cal D}^-_{\a\da} \bar\chi^{\da}\,,  \label{rescL}
\end{align}
where
$L_{\rm YM}$ \p{Lexpl}, the field strength $F$ and the covariant derivatives ${\cal D}$ do not depend on the coupling constant $g$. It is only present in \p{rescL} as an overall factor in front of $L_{\rm YM}$. We use the rescaled Lagrangian \p{rescL} in the path integral representation of the correlation function, 
\begin{align}
\vev{J_{\mu}(x_1)\ldots J_{\nu}(x_n)} = \int D \Phi \, e^{i \int d^4 x L(x)} J_{\mu}(x_1)\ldots J_{\nu}(x_n)\,,  \label{path-int}
\end{align} 
where $D\Phi$ denotes the integration measure for all the fields of the theory.   
Differentiating both sides of \p{path-int} with respect to $g^2$ we obtain
\begin{align}
& g^2 \frac{\pa}{\pa g^2} \vev{J_{\mu}(x_1)\ldots J_{\nu}(x_n)} = 
-\frac{i}{g^2} \int D \Phi \, e^{i \int d^4 x L(x)} \int d^4 x\, L_{\rm YM}(x) J_{\mu}(x_1)\ldots J_{\nu}(x_n) \notag\\
& = -\frac{i}{g^2} \int d^4 x\, \vev{L_{\rm YM}(x) J_{\mu}(x_1)\ldots J_{\nu}(x_n)}\,.
\end{align}
In this way we express the one-loop correction of the correlation function of currents \p{pert-exp} in terms of a correlator with one additional Lagrangian point, calculated at the lowest perturbative order (Born level),
\begin{align}
g^2 G^{\text{1-loop}}_{\mu\ldots\nu}(x_1,\ldots,x_n) = -i \int d^4 x \, \vev{L_{\rm YM}(x) J_{\mu}(x_1)\ldots J_{\nu}(x_n)}_{\rm Born} \,.  \label{Born}
\end{align}
The calculation of the $n-$point correlator of currents is done in two steps:
\begin{itemize}
\item Calculation of the $(n+1)-$point correlator with the operator $L_{\rm YM}$ at the insertion point at  Born level,
\begin{align}\label{e210}
G_{\mu \ldots \nu} (x_1,\ldots,x_n|x) =\vev{L_{\rm YM}(x) J_{\mu}(x_1)\ldots J_{\nu}(x_n)}_{\rm Born}\,.
\end{align}
As we will see later on, this correlator is a rational function. 
\item Space-time integration over the Lagrangian insertion point, 
\begin{align}
g^2 G^{\text{1-loop}}_{\mu\ldots\nu}(x_1,\ldots,x_n) = -i \int d^4 x \, G_{\mu \ldots \nu} (x_1,\ldots,x_n|x)\,.  \label{Born->loop}
\end{align}
\end{itemize} 

In the following subsections we proceed by successively implementing both tasks. As explained earlier, at this perturbative level the conformal symmetry of the Lagrangian \p{rescL} is preserved. Consequently, both the rational function in \p{e210} and its integral in \p{Born->loop} enjoy manifest conformal covariance. This circumstance greatly facilitates the further steps in evaluating the charge-charge correlation.

\subsection{Feynman diagrams at order $O(g^2)$}

\begin{figure}
\begin{center}
\begin{tabular}{ccc}
\includegraphics[width=4cm]{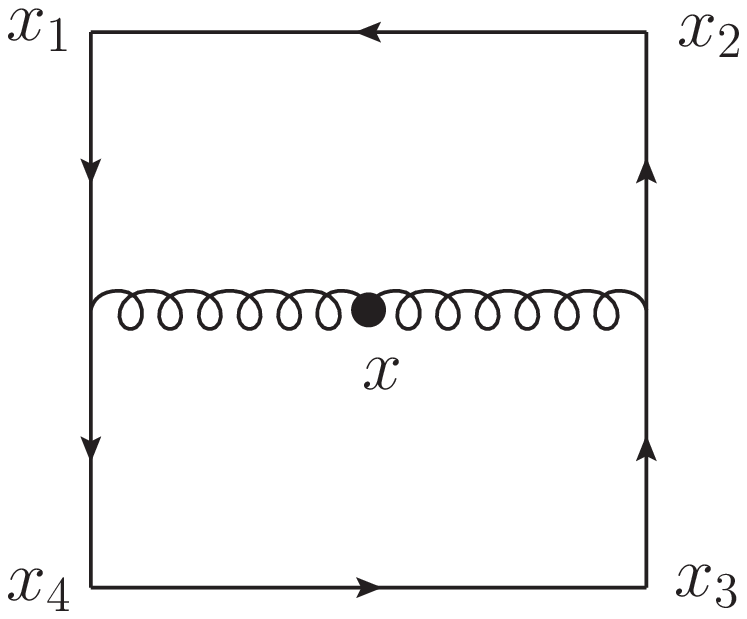} & 
\includegraphics[width=4cm]{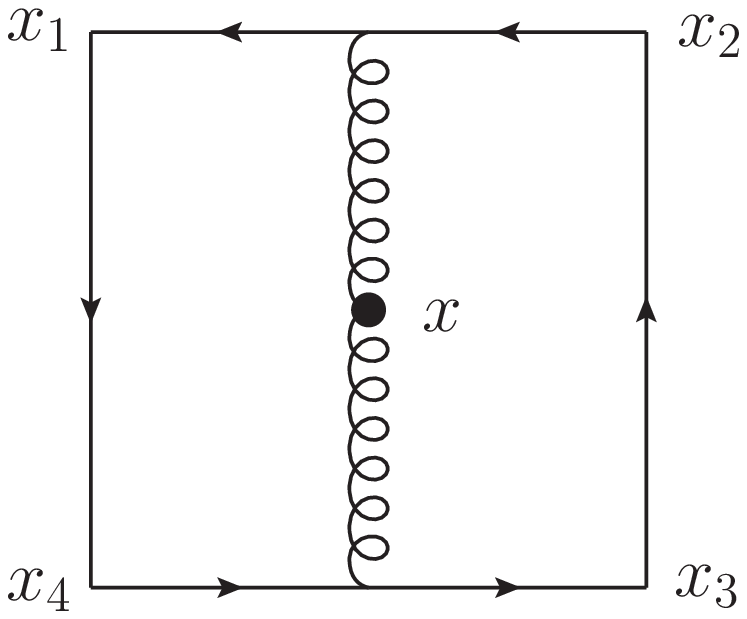} &
\includegraphics[width=4cm]{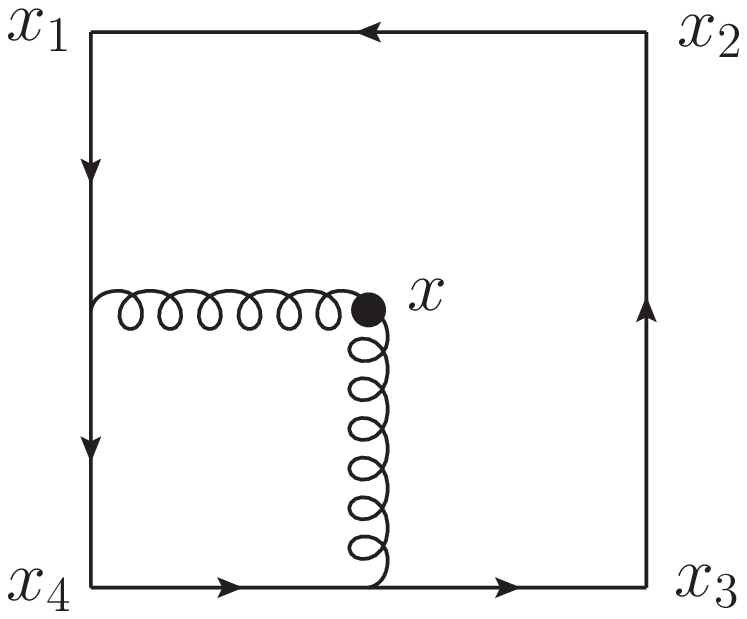} \\
\includegraphics[width=4cm]{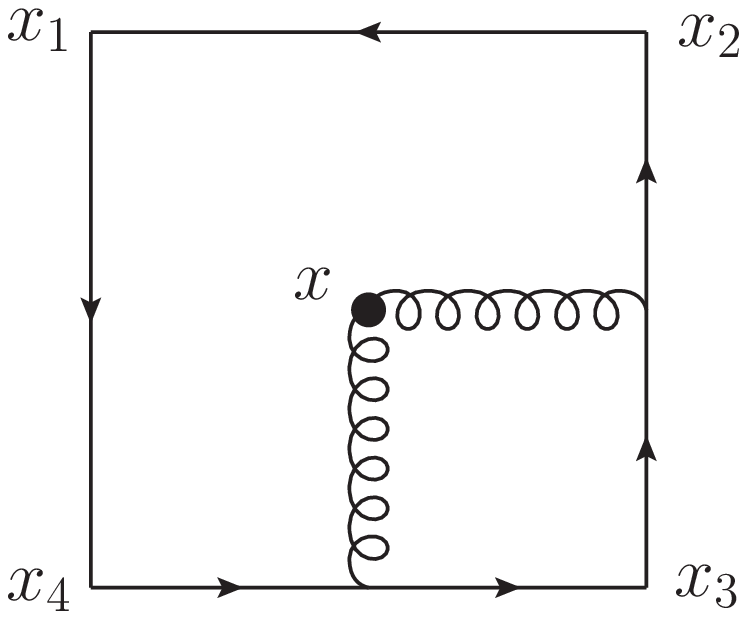} & 
\includegraphics[width=4cm]{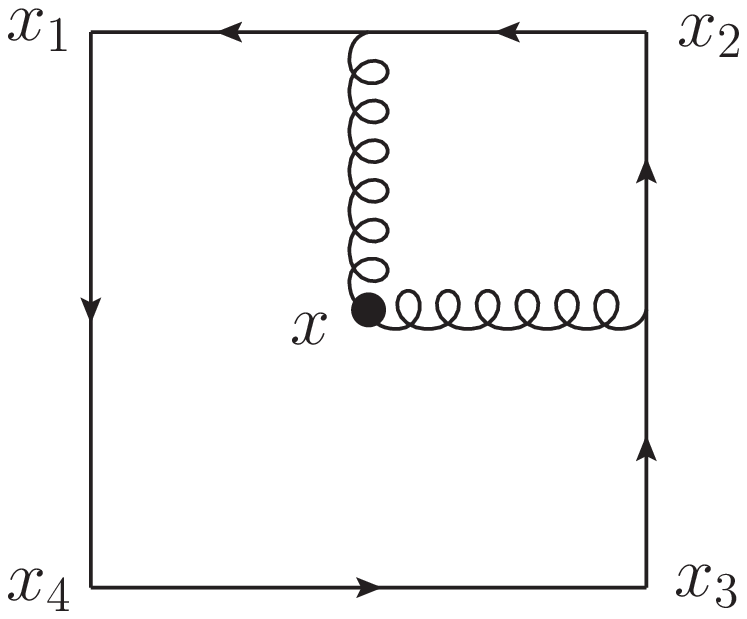}&
\includegraphics[width=4cm]{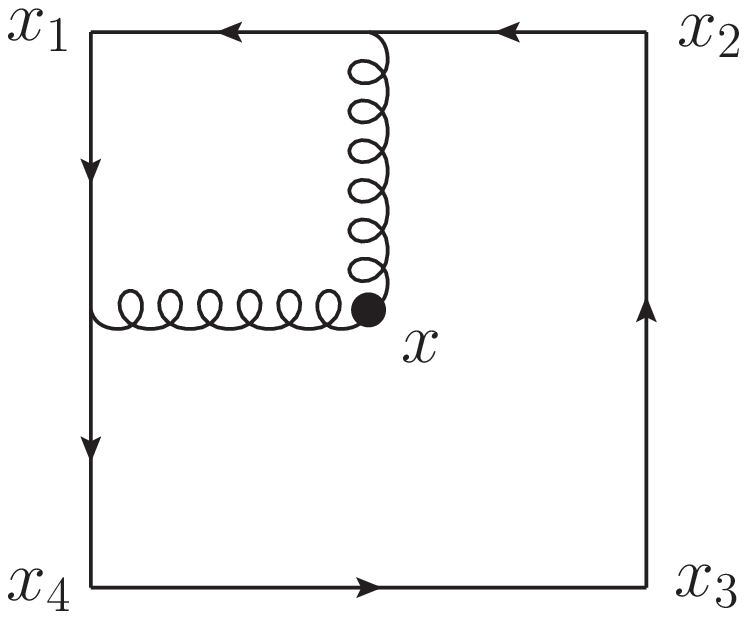}
\end{tabular}
\end{center}
\caption{Feynman diagrams contributing to $\vev{L_{\rm YM}(x) J(x_1)J(x_2)J(x_3)J(x_4)}_{\rm Born}$ built out of fermion propagators and T-blocks shown in Fig.~\ref{blocks}.A.}
\label{integrand1}
\end{figure}

The tree-level correlator of currents $J_{\a\da} = \psi_{\a}\bar\psi_{\da}$ is a sum of products of fermion propagators. For example in the four-point case we obtain
\begin{align}
\vev{J_{\a\da}(x_1) J_{\b\db}(x_2) J_{\gamma\dot\gamma}(x_3) J_{\delta\dot\delta}(x_4)}_{\text{tree}}= -\frac{N_c}{(2\pi^2)^4} \sum_{\sigma\in S_4/\mathbb{Z}_4} \frac{(x_{\sigma_1\sigma_2})_{\a\db}(x_{\sigma_2\sigma_3})_{\b\dot\gamma} (x_{\sigma_3\sigma_4})_{\gamma\dot\delta} (x_{\sigma_4\sigma_1})_{\delta\dot\alpha}}{x_{\sigma_1\sigma_2}^4 x_{\sigma_2\sigma_3}^4 x_{\sigma_3\sigma_4}^4 x_{\sigma_1\sigma_4}^4} 
\label{tree-level}
\end{align}
where the sum over the inequivalent permutations amounts to Bose symmetrization. Here we tacitly imply permuting the Lorentz indices along with the points $x_i$. Expression \p{tree-level} is manifestly conformal, carrying conformal weight $(+3)$ at each point.

\begin{figure}
\begin{center}
\begin{tabular}{cccc}
\includegraphics[width=3.7cm]{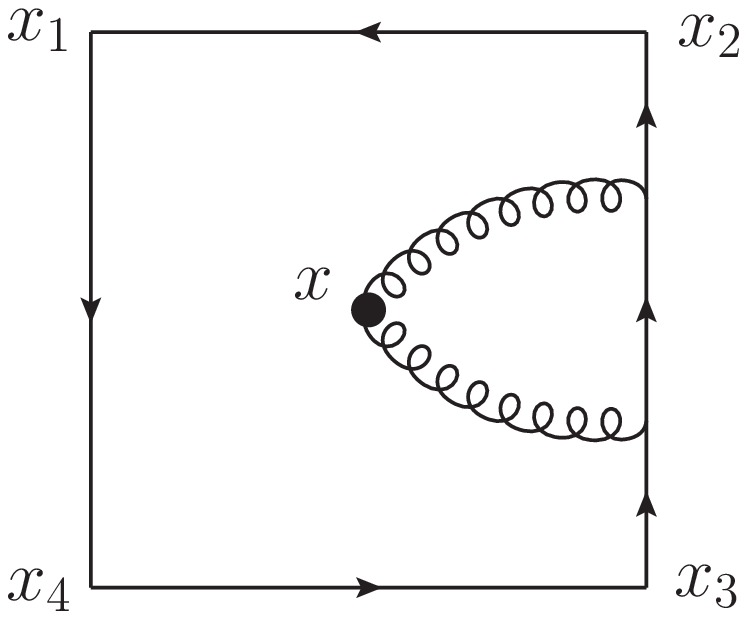} & 
\includegraphics[width=3.7cm]{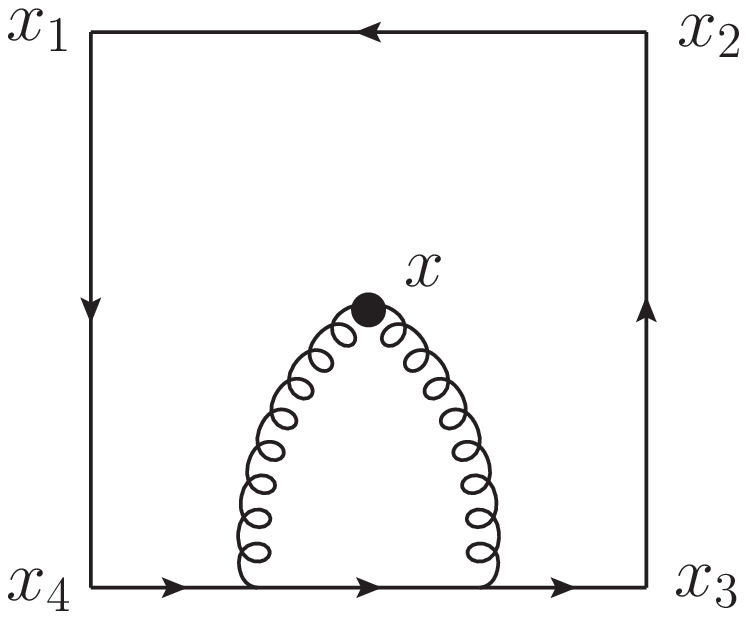} &
\includegraphics[width=3.7cm]{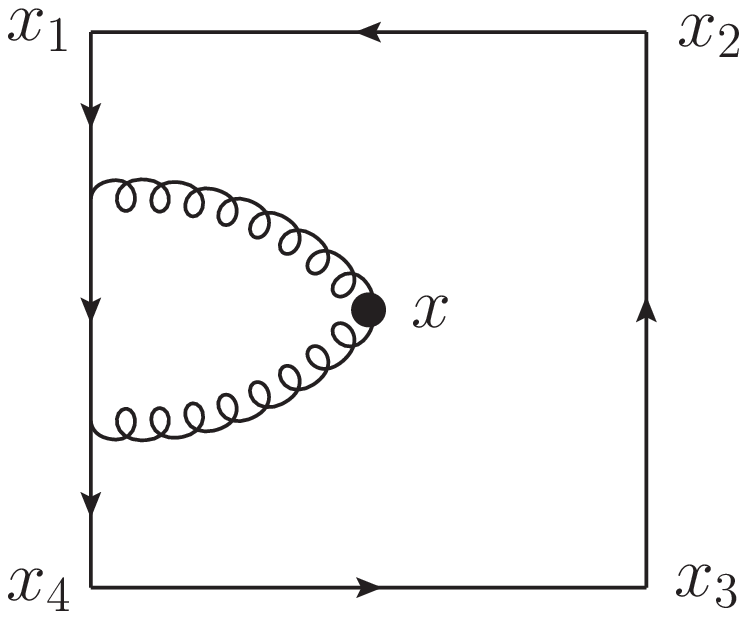} &
\includegraphics[width=3.7cm]{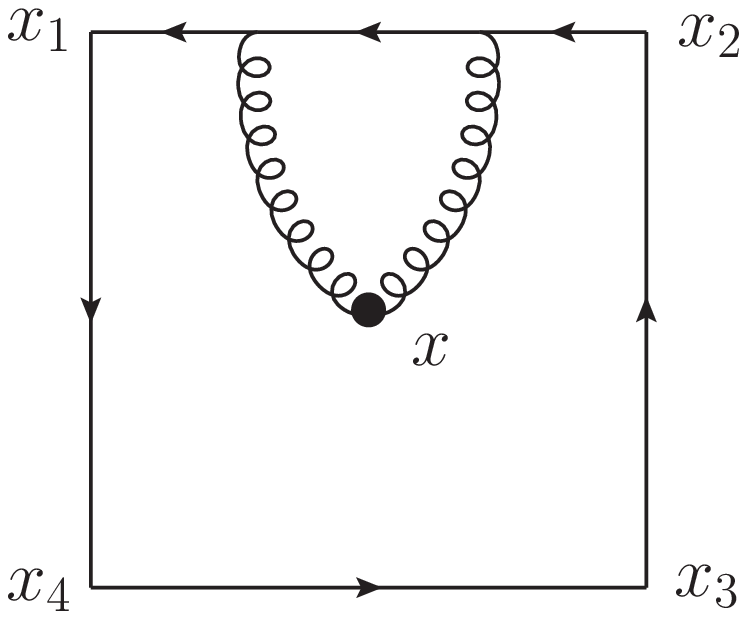}
\end{tabular}
\end{center}
\caption{Feynman diagrams contributing to $\vev{L_{\rm YM}(x) J(x_1)J(x_2)J(x_3)J(x_4)}_{\rm Born}$  built out of fermion propagators and  Lagrangian insertions of  the type shown in Fig.~\ref{blocks}.B.}
\label{integrand2}
\end{figure}

Let us now consider the Born-level correlator of  four currents with one Lagrangian insertion. The corresponding Feynman diagrams are shown in Figs.~\ref{integrand1} and \ref{integrand2}. We also need to Bose symmetrize them by adding five noncyclic permutations of the external points like at tree level. Each diagram contains two interaction vertices $g \psi A \bar\psi$ from the Lagrangian \p{Lexpl}, so the Born-level correlator is of order $O(g^2)$ in the coupling. 

\begin{figure}
\begin{center}
\begin{tabular}{ccc}
\includegraphics[width=5cm]{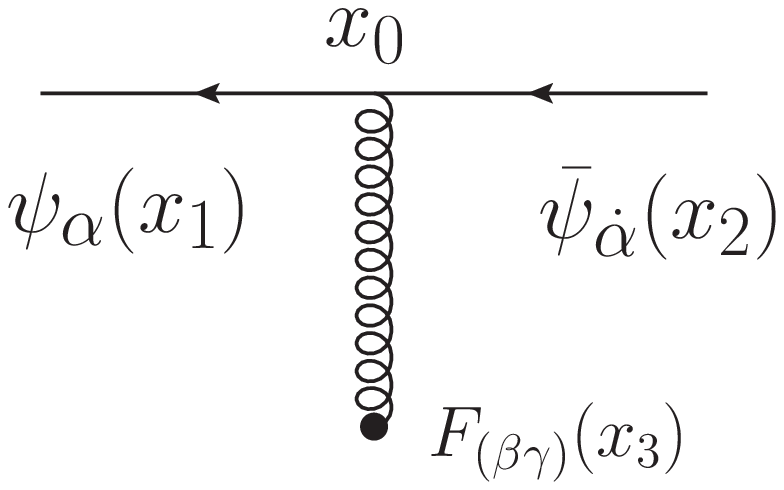} & \qquad\qquad\qquad & \includegraphics[width=6cm]{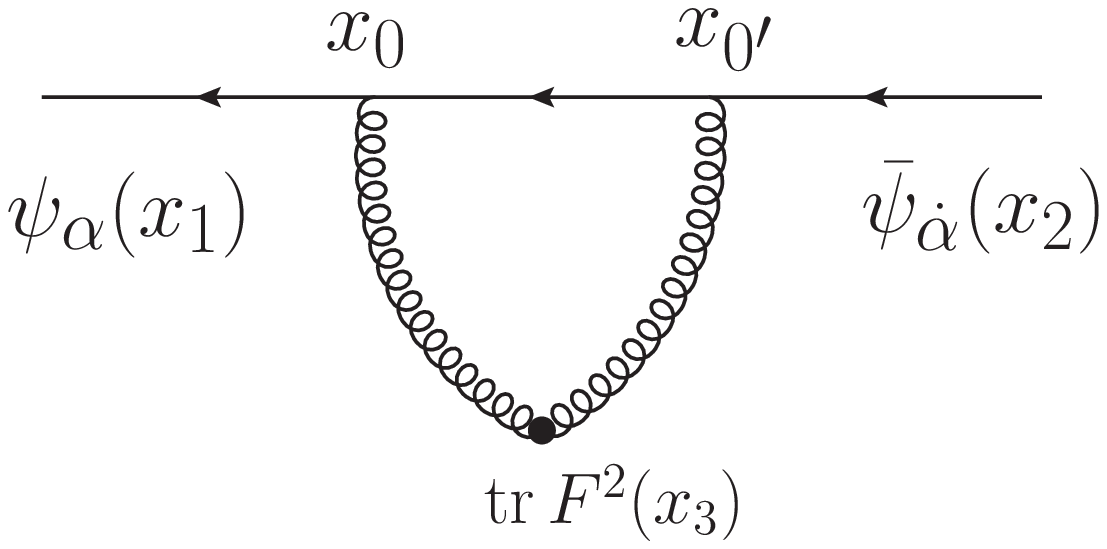} \\
(A) & & (B) 
\end{tabular}
\end{center}
\caption{(A) T-block diagram \p{T-block} representing $\vev{\psi \bar\psi F}$ at Born level. (B) Lagrangian insertion in the fermion propagator \p{propCorr}.}
\label{blocks}
\end{figure}

Inspecting the Feynman diagrams in Fig.~\ref{integrand1} we see that they are products of fermion propagators and the so-called T-blocks -- an interaction vertex of two fermions and a gluon with the field strength at the gluon propagator end, see Fig.~\ref{blocks}.A,   
\begin{align}
&\vev{\psi_\a(x_1) \bar\psi_\da(x_2) F_{a}^{\mu\nu}(x_3)}_{g}  = \frac{i g\, T_a}{(2\pi)^6}
\int d^4 x_0 \, \pa_{\a}^{\dot\beta}\frac{1}{x_{10}^2}  \pa^{\beta}_{\da} \frac{1}{x_{20}^2}
\sigma^{[\nu}_{\b\db}
\pa^{\mu]}\frac{1}{x_{30}^2} \notag\\
&= \frac{g\, T_a}{(2\pi)^4}\left\{ \frac{4(x_{12})_{\a\da}}{x^4_{12}} \frac{x_{31}^{[\mu} x_{32}^{\nu]}}{x^2_{13} x^2_{23}} - \frac{(x_{12} \tx_{23} \sigma^{[\mu} \tilde\sigma^{\nu]} x_{32})_{\a\da}}{2x^2_{12} x^2_{13} x^4_{23}}
- \frac{(\tx_{21} x_{13} \tilde\sigma^{[\mu} \sigma^{\nu]} \tx_{31})_{\a\da}}{2x^2_{12} x^4_{13} x^2_{23}} \right\}\,.
\label{T-block}
\end{align}
 The integration in eq.~\p{T-block} is carried out by means of the star-triangle relation, which yields a rational function. This object is gauge covariant (not invariant), consequently it is not conformally covariant.

The diagrams in Fig.~\ref{integrand2} contain another building block depicted in Fig.~\ref{blocks}.B. It is a Born-level diagram representing the coupling of  the Yang-Mills Lagrangian to the fermion propagator. This diagram contains two interaction vertices with integrations $\int d^4 x_0 d^4 x_{0'}$. Remarkably, it turns out to be a rational function,\footnote{This result is far from obvious. It has been obtained in two independent ways.  Firstly, we split the two field strengths $F(x_3)F(x_{3'})$ apart, then evaluated the new four-point function and took the limit $x_{3'}\to x_3$ in a symmetric way. The second method used the knowledge that the two-point functions of the super-descendants of the half-BPS operator $\tr(\phi^2)$ in $\cN=4$ sYM are protected. The corresponding one-loop diagrams with Lagrangian insertion are made from several subdiagrams, one of them being \p{propCorr},  the rest are explicitly  known. This allowed us to express the unknown building block  as a sum of known ones.  }
\begin{align}
&\vev{\psi_\a(x_1) \bar\psi_\da(x_2)\, \tr \, (F_{\mu\nu}^2)(x_3)}_{g^2}  \notag\\
& = \frac{2ig^2 C_F}{(2\pi)^{10}} \int d^4 x_0 d^4 x_{0'} \pa_{\a\db} \frac{1}{x_{10}^2} \pa_{\b\dot\gamma} \frac{1}{x_{00'}^2} \pa_{\gamma \da} \frac{1}{x_{20'}^2} \left[2\epsilon^{\b\gamma}\epsilon^{\db\dot\gamma} \pa_\mu \frac{1}{x_{30}^2} \pa^\mu \frac{1}{x_{30'}^2} 
-\pa^{\dot\gamma\gamma} \frac{1}{x_{30}^2} \pa^{\db \beta} \frac{1}{x_{30'}^2}  
\right] \notag\\
&= \frac{2ig^2 C_F}{(2\pi)^6} 
\frac{(x_{12})_{\a\da}}{x_{12}^4 x_{13}^4 x_{23}^4}  [x_{12}^4 + x_{13}^4+x_{23}^4 - 2 x_{12}^2 x_{13}^2 - 2 x_{12}^2 x_{23}^2 - 2 x_{13}^2 x_{23}^2]\,, 
\label{propCorr}
\end{align}
where
\begin{align}\label{}
C_F =\frac{N_c^2-1}{2 N_c}
\end{align} 
 is the Casimir invariant of the fundamental color representation of the fermions. Similarly to \p{T-block},  expression \p{propCorr} is not conformally covariant.

Combining all the ingredients above, we obtain  a rather bulky expression for the $(4+1)$-point Born-level correlation function, which is the one-loop integrand of the correlator of four currents, see eq.~\p{Born->loop}. Let us emphasize once more that the integrand is a finite rational function living in four space-time dimensions. Despite of the fact that the individual Feynman graphs contributing to the integrand are not conformally covariant, the integrand itself (i.e. the gauge invariant sum of all the graphs) is conformal. We explicitly checked that this property holds.

\begin{figure}
\begin{center}
\begin{tabular}{ccccc}
$\begin{array}{c}\includegraphics[width=4cm]{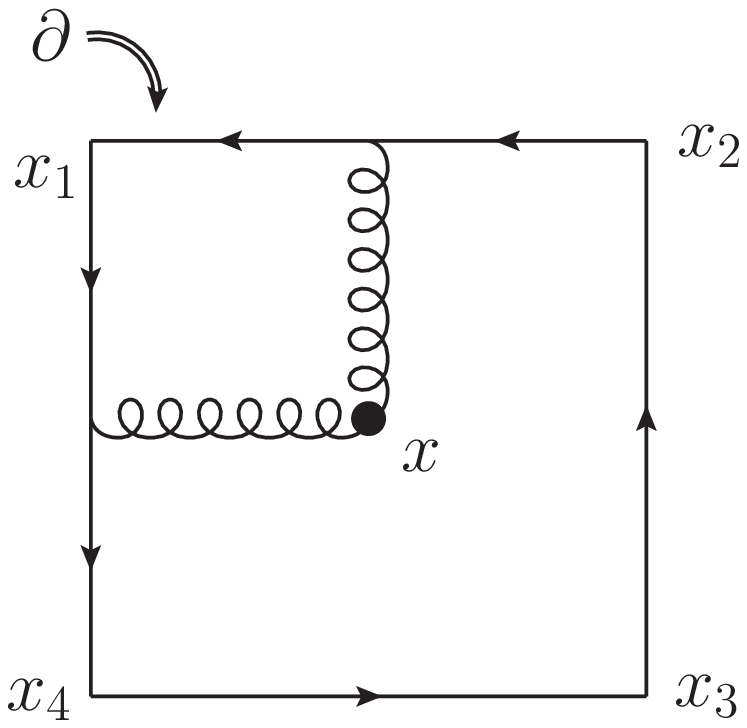}
\end{array}$ & 
$\Longrightarrow$ & 
$\begin{array}{c}\includegraphics[width=4cm]{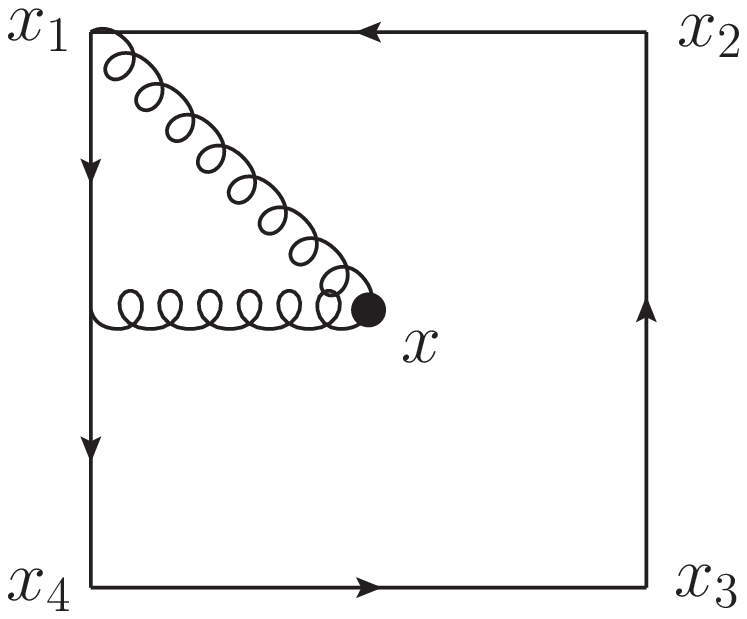}
\end{array}$ & $\Longleftarrow$ &
$\begin{array}{c}\includegraphics[width=4cm]{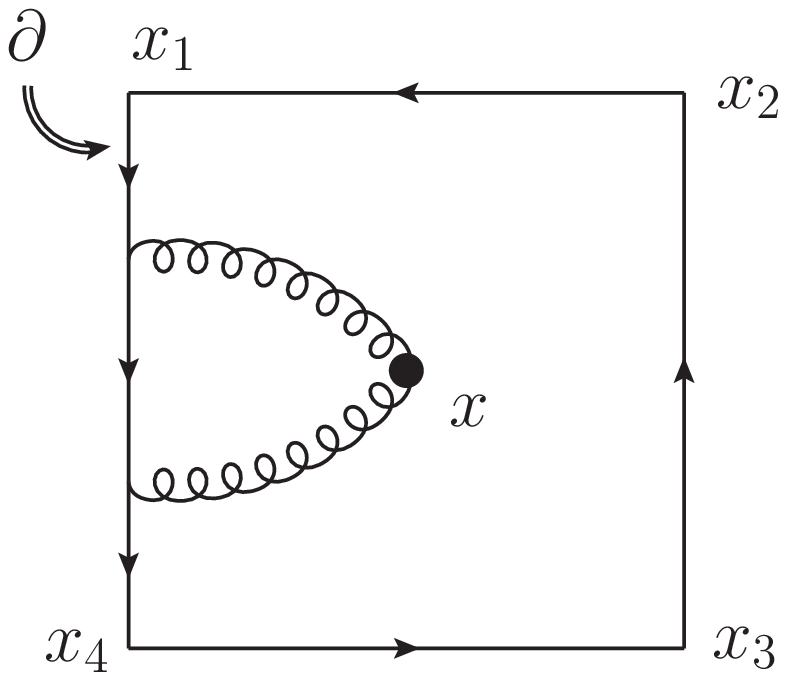}
\end{array}$
\end{tabular}
\end{center}
\caption{The current conservation Ward identity \p{WardCC} can be easily verified at the level of the correlator Feynman diagrams given in Figs.~\ref{integrand1} and \ref{integrand2}.}
\label{conservation}
\end{figure}

An additional check of our result for  the integrand is provided by current conservation. The divergence of the correlator with respect to each current point must vanish (up to possible contact terms), e.g.
\begin{align}
\pa^{\da\a}_1 \vev{L_{\rm YM}(x) \, J_{\a\da}(x_1) \ldots J_{\delta\dot\delta}(x_4)}_{\rm Born} = 0 \,. \label{WardCC}
\end{align} 
We  checked that the integrand obtained has this property.
The current conservation can be easily seen at the level of the Feynman graphs. Since by definition
\begin{align}
\pa^{\da\a} \vev{\psi_{\a}(x)\bar\psi_{\db}(0)} = \delta^{\da}_{\db}\delta^{(4)}(x)\,,  \label{Dfermprop}
\end{align}
the derivative shrinks the fermion propagator  to a point, and the `shrunk' Feynman diagrams cancel pairwise, as depicted in Fig.~\ref{conservation}. Using eq.~\p{Dfermprop} we can derive first order differential equations for the building blocks \p{T-block} and \p{propCorr}. We  checked that the provided rational expressions satisfy these differential equations.

\subsection{Result for the four-point correlation function }

Having obtained the one-loop integrand, at the next step we  integrate over the Lagrangian insertion point, see eq.~\p{Born->loop}. 
Even though the correlator of four currents is UV finite in the one-loop approximation, the integration of some individual terms of the integrand could lead to UV divergences. So we need to introduce an intermediate regulator. We use dimensional regularization and integrate over the Lagrangian point in $D = 4-2\ep$ dimensions. Since the integrand carries  Lorentz spinor indices, we have to specify how to treat this objects in $D$ dimensions. We assume that the $\sigma$ matrices are four-dimensional and we do the spinor algebra in four space-time dimensions. When rewriting the integrand numerator we have the choice between two prescriptions: either $(x \tilde x)_{\a}{}^{\b} = \delta_{\a}^{\b} x^2$ or $(x \tilde x)_{\a}{}^{\b} = \delta_{\a}^{\b} (x^2-\mu^2)$ where $x^2$ is $D$-dimensional and $\mu^2$ is the $(-2\epsilon)$-dimensional part of $x^2$ \cite{Bern:1995db}. We did the calculations in both schemes and found identical results. 

The integrals that we need can be mapped to the well-known families of one-loop amplitude integrals with massive legs and massless internal propagators. The numerators of the integrands are tensors of maximal rank two. We perform the tensor reduction procedure in coordinate space in order to reduce them to scalar one-loop integrals (see Appendix~\ref{tens-red}). Then we do IBP reduction of the scalar integrals to a set of master integrals: bubble, three-mass triangle, and four-mass box integrals. Removing the regulator $\epsilon \to 0$ we find that the triangle integrals are suppressed, and all the $\epsilon$-poles in the individual terms cancel out in the sum. 
The logarithmic terms $\log(x_{ij}^2)$ in the $\epsilon$-expansion of the bubble integrals combine together into logarithms of the conformal cross-ratios 
\begin{align}\label{e2.17}
u = \frac{x_{12}^2 x_{34}^2}{x_{13}^2 x_{24}^2} \;,\qquad
v = \frac{x_{14}^2 x_{23}^2}{x_{13}^2 x_{24}^2} 
\;.
\end{align}
Thus the one-loop correlator takes the following form  
\begin{align}
&G^{\text{1-loop}}(x_1,\ldots,x_4)\nt
&= -\frac{4N_c C_F}{(2\pi)^{10}}\left[ R_{c}(x)\,  \Phi^{(1)}(u,v) + R_u(x)\, \log(u) + R_v(x)\, \log(v) + R_r(x)\right] + R'_r(x) \,, \label{JJJJ}
\end{align}
where we omit the Lorentz indices. The various $R$'s are rational functions of the coordinates. They are  conformally covariant  tensors carrying spin 1 and weight $(+3)$ at each point. The function $\Phi^{(1)}$ is the well-known cross integral (four-mass box integral in the amplitude literature),
\begin{align}
\Phi^{(1)}(u,v) = \frac{1}{\bar z- z}\left[ 2{\rm Li}_2\left(\frac{z}{z-1}\right) - 2{\rm Li}_2\left(\frac{\bar z}{\bar z-1}\right) - \log \left( \frac{z \bar z}{(1-z)(1-\bar z)} \right) \log \left( \frac{1-z}{1-\bar z} \right) \right] \label{Phi1}
\end{align} 
given here in the $z,\bar z$ parametrization of the conformal cross-ratios 
\begin{align}\label{defzz}
u = z \bar z \;,\qquad
v = (1-z)(1-\bar z)\,.
\end{align}
We remark that $\Phi^{(1)}$ is regular at $z = \bar z$. The rational term $R_r(x)$ in \p{JJJJ} comes from the $\ep \cdot \frac{1}{\ep} = 1$ cancellations between the simple poles of the bubbles and the $D$-dependent coefficients of the IBP relations and the tensor reductions. 

The conformal covariance of our final result \p{JJJJ} is a strong check of our calculation. 
Note however that the rational functions $R_c,\,R_u,\,R_v,\, R_r$ do not arise in a manifestly conformal form in our Feynman graph calculation. To find a  conformal representation for them we study their singularities, construct a conformal ansatz for their numerators, and match them with the ansatz. In the case of interest we need to list the conformal polynomial tensors of conformal weight $(-6)$ at each point. This is done as follows.
We use the conformal expressions $(x_{ij})_{\a_i \da_j}$, $(x_{i k}\tilde x_{kj})_{\a_i \a_j}$, and $(\tilde x_{i k}x_{kj})_{\da_i \da_j}$ as elementary constituents, and form products that are conformal tensors carrying the eight Lorentz spinor indices of the currents, $\a\da\ldots\delta\dot\delta$. In this way we find 242 linearly independent over $\mathbb{Q}$ polynomial conformal tensors. Then we complement them with polynomials in $x_{ij}^2$ such that each tensor carries  conformal weight $(-6)$ at each point, and we find that only 1024 of them are linearly independent over $\mathbb{Q}$. We take them as a basis of conformal structures for the numerators of $R_c,\,R_u,\,R_v,\, R_r$.

The last rational term $R'_r(x)$ in \p{JJJJ} stays apart. It comes from integrating out the Lagrangian insertion point in the Feynman diagrams in Fig.~\ref{integrand2}. Integrating the building block \p{propCorr}  we find the one-loop propagator correction  
\begin{align}\label{2.22}
\int d^4 x_3\, \vev{\psi_\a(x_1) \bar\psi_\da(x_2)\, \tr \, (F_{\mu\nu}^2)(x_3)}_{g^2}  = -\frac{3g^2 C_F}{(2\pi)^4}\frac{(x_{12})_{\a\da}}{x_{12}^4}\,,
\end{align}
which is proportional to the free fermion propagator.\footnote{We recall that the one-loop correction to the fermion propagator is scheme and gauge dependent. } Consequently, $R'_{r}$ is proportional to the tree-level correlator \p{tree-level}, 
\begin{align}
R'_r(x_1,x_2,x_3,x_4) =
\frac{3 g^2 N_c C_F}{8(2\pi^2)^5} \sum_{\sigma\in S_4/\mathbb{Z}_4} \frac{(x_{\sigma_1\sigma_2})_{\a\db}(x_{\sigma_2\sigma_3})_{\b\dot\gamma} (x_{\sigma_3\sigma_4})_{\gamma\dot\delta} (x_{\sigma_4\sigma_1})_{\delta\dot\alpha}}{x_{\sigma_1\sigma_2}^4 x_{\sigma_2\sigma_3}^4 x_{\sigma_3\sigma_4}^4 x_{\sigma_1\sigma_4}^4} \,.
\end{align}  
Obviously, this term is conformal on its own and satisfies the current conservation.

Differentiating $\Phi^{(1)}$ with respect to the variables $u,v$ gives
\begin{align}
& \pa_u \Phi^{(1)} = \frac{1}{\la^2} \left[ (1-u+v)  \Phi^{(1)} + 2\log u + \frac{1-u-v}{u} \log v \right], \notag\\
& \pa_v \Phi^{(1)} = \frac{1}{\la^2} \left[ (1+u-v)  \Phi^{(1)} + 2\log v + \frac{1-u-v}{v} \log u \right], \label{Dphi}
\end{align}
where $\la^2(u,v)  = 1 +u^2 +v^2 - 2 u -2 v - 2 u v= (\bar z- z)^2$.
Thus one sees that the current conservation Ward identity for \p{JJJJ} relies on a nontrivial system of differential equations relating the functions $R_c,\,R_u,\,R_v,\, R_r$. We explicitly checked that \p{JJJJ} does satisfy current conservation with respect to each external point. 

An essential drawback of the representation \p{JJJJ} of our result  is that the 
rational functions $R_c,\,R_u,\,R_v,\, R_r$ contain spurious poles at $z = \bar z$. We would like to obtain another representation for the correlator which has only the physical singularities determined by the OPE. By 
differentiating further eqs.~\p{Dphi} we obtain relations which allow us to express  $\log(u)$, $\log(v)$, $\Phi^{(1)}$   in terms of  the second-order derivatives $\pa^2_{uu} \Phi^{(1)}$, $\pa^2_{vv} \Phi^{(1)}$, $\pa^2_{uv} \Phi^{(1)}$.   Substituting these expressions back in \p{JJJJ} we find that all the $(z-\bar z)$ factors in the denominators of the rational coefficients disappear. The resulting expression contains only physical singularities.
Finally, we assemble the $x_{ij}^2$ into conformal cross-ratios and the correlator takes the following form
\begin{align}
G^{\text{1-loop}}(x_1,\ldots,x_4) =&  -\frac{4 N_c C_F}{(2\pi)^{10}} \frac{1}{(x_{13}^2 x_{24}^2)^4}\biggl[ \pa_{uu}^2 \Phi^{(1)} \sum_{I} c_{uu}^{(I)}(u,v) T_I  + \pa_{vv}^2 \Phi^{(1)} \sum_{I}c_{vv}^{(I)}(u,v) T_I \notag\\
& + \pa_{uv}^2 \Phi^{(1)} \sum_{I} c_{uv}^{(I)}(u,v) T_I  + \sum_{I} c_{0}^{(I)}(u,v) T_I \biggr] +  R'_r  \,. \label{JJJJ2}
\end{align}
The coefficients $c(u,v)$ are polynomial in $u,v,u^{-1},v^{-1}$. The Lorentz indices are carried by 31 conformal structures $\{ T_{I}\}_{I=1}^{31}$ with conformal weight $(-1)$ at each point.\footnote{It would be interesting to compare this number to the known basis of 43 conformal tensors in the most general correlator of four spin-one vectors, see e.g. \cite{Dymarsky:2013wla,DK}.} We can write them down schematically in the following form 
\begin{align}
T_I = \sum_{J} (b^{(J)}_0 + b^{(J)}_u u + b_v^{(J)} v) \, t_J\,,
\end{align}
where $b_0,\,b_u,\,b_v$ are integers and the tensors $t_J$ are formed from conformally covariant products of $(x_{ij})_{\a_i\dot\a_j}$, $(x_{i k}\tilde x_{kj})_{\a_i \a_j}$ and $(\tilde x_{i k}x_{kj})_{\da_i \da_j}$.\footnote{We need 9 products of four  $(x_{ij})_{\a_i\dot\a_j}$, e.g. $(x_{13})_{\a\dot\gamma} (x_{21})_{\b\da} (x_{34})_{\gamma\dot\delta} (x_{42})_{\delta\dot\beta}$, and 45 products of two $(x_{ij})_{\a_i\dot\a_j}$ and $(x_{i k}\tilde x_{kj})_{\a_i \a_j}$ and $(\tilde x_{i k}x_{kj})_{\da_i \da_j}$, e.g. $\frac{1}{x_{14}^2} (x_{13})_{\a\dot\gamma} (x_{41})_{\delta\da} (\tilde x_{21}x_{14})_{\db\dot\delta} (x_{24} \tilde x_{43})_{\b\gamma}$.} The tensors $T_I$ are parity even and real in Minkowski kinematics, i.e. 
converting the spinor indices to vector indices, 
\begin{align}
(T_I)_{\a\da \ldots \delta\dot\delta} \; \tilde\sigma^{\da\a}_{\mu}\, 
\tilde\sigma^{\db\b}_{\nu}
\tilde\sigma^{\dot\gamma\gamma}_{\la}
\tilde\sigma^{\dot\delta\delta}_{\rho}\,,
\end{align}
we find that the resulting expressions do not involve the Levi-Civita tensor.  The explicit expressions for $T_I$ and $c(u,v)$ are provided in the ancillary file. Of course, our choice of basis of conformal tensors in \p{JJJJ2} is not unique. The main advantage of this choice is that the basis tensors and their coefficients do not have unphysical singularities.

We calculated the correlation function \p{JJJJ2} of four currents $J_{\a\da}  = \psi_{\a} \bar\psi_{\da}$ built from one Weyl spinor and its conjugate (or equivalently, from one Majorana fermion). One can easily see that the one-loop Feynman rules yield the same result for the correlator of four currents $\tilde{J}_{\a\da}  = \chi_{\a} \bar\chi_{\da}$. Since the interactions do not mix the $\psi$ and $\chi$ fields, see \p{Lexpl}, the connected components of the correlators involving a mixture of $J$ and $\tilde J$ have to vanish. We conclude that the correlator of four vector electromagnetic currents $J^{(v)}$ from  \p{currentsdef} equals twice \p{JJJJ2}. If instead we use an odd number of axial currents from  \p{currentsdef}, the one-loop correlator vanishes.

\section{From the correlation function to the charge-charge correlation}\label{s4}

In this section we evaluate the event shape function \p{zetadef} for the charge-charge correlation at one loop. We substitute our result \p{JJJJ2} for the correlation function of  four electromagnetic  currents $J \equiv J^{(v)}$ in the definitions \p{124} and \p{E-new}. The charge flow correlation  between two detectors in the direction of $\vec{n} $ and $ \vec{n}'$ is given by
\begin{align} \label{e3.1}
&\vev{\mathcal{Q}(\vec{n}) \mathcal{Q}(\vec{n}') }  =
\sigma_\text{tot}^{-1}    \int d^{4} x_{14} e^{i q x_{14} }  \int_{-\infty}^\infty d x_{2-} d x_{3-}     \\
&  \lim_{ x_{2+}, x_{3+} \rightarrow \infty }   x_{2+}^2 x_{3+}^2
 \vev{  J_{\a \da} (x_1) \, J_{\b \db } ( x_2  )  \,  J_{\g \dg}( x_3 ) \, J_{\d \dd} (x_4)    }\, \ep^{\a\delta} \ep^{\da\dot\delta}  \, \bar{n}^{\db \b } \bar{n}'^{\dg \g } \,.       \nn
\end{align} 
As explained in Sect.~\ref{s1.4}, we consider a rotationally invariant event shape, which allows us to contract the Lorentz indices at points 1 and 4.  

\subsection{Detector limit}\label{s3.1}

In \p{e3.1} the two detectors at points 2 and 3  are infinitely far away from the sink and source at points 1 and 4. 
 In order to take the detector limit, we introduce the light-cone parametrization \p{x} of the detectors,
\begin{align}
x_2^\mu = x_{2+} \, n^\mu + x_{2-} \, \bar{n}^\mu \;,\qquad 
x_3^\mu = x_{3+} \, n'^\mu + x_{3-} \, \bar{n}'^\mu   \,,
\end{align} 
in the basis of the lightlike vectors $n^2 = \bar n^2 = ( n' )^2 = (\bar n')^2 = 0$. 
The detector currents are  projected, $J_+(x_2) = \bar n^\mu J_\mu(x_2)$ and  $J_+(x_3) = \bar n'^\mu J_\mu(x_3)$.  
Sending the detectors to infinity amounts to the limit $x_{2+} , x_{3+} \to \infty$ in which
\begin{align} 
\label{eq:3.3}
& x_{12}^2 \rightarrow  \,  2 x_{2+} x_{21-} ,  \; \; x_{13}^2 \rightarrow   \, 2 x_{3+} x_{31-}  , \; \; 
x_{24}^2 \rightarrow  \,2 x_{2+}  x_{24-} ,  \; \; x_{34}^2 \rightarrow   \, 2  x_{3+}  x_{34-} , \nn \\
& x_{14}^2 \rightarrow  x_{14}^2, \; \;  x_{23}^2 \rightarrow  -2 ( n n' ) x_{2+} x_{3+} \,.
\end{align} 
Here we use the short-hand notations
\begin{align}\label{e34}
& x_{21-} = x_{2-} (n\bar n) - (n x_1) \;,\; 
x_{24-} = x_{2-} (n\bar n) - (n x_4) \;,\notag\\
& x_{31-} = x_{3-} (n'\bar n') - (n' x_1) \;,\; 
x_{34-} = x_{3-} (n'\bar n') - (n' x_4)\,. 
\end{align}

We find that the detector limit  takes the following form\footnote{In taking the limit we ignore the rational term $R'_r$ in \p{JJJJ2}. It is proportional to the Born-level correlator and the corresponding event shape is a contact term \cite{Belitsky:2013bja}.}
\begin{align} \label{3.5}
&  \lim_{ x_{2+},x_{3+} \rightarrow \infty }  x_{2+}^2 x_{3+}^2 \, \vev{J^\mu (x_1) J_+ (x_2) J_+ (x_3)  J_\mu (x_4) }
 =-\frac{g^2 N_c C_F}{2^4 \pi^{10}}\frac{(n  \bar{n} ) ( n'   \bar{n}')  }{x_{14}^8  \, (n  n')^3 }  \, \mathcal{G}(u,v ,  \g ) \,,
  \end{align}  
  where (see \p{e2.17})
\begin{align}\label{3.6}
u = \frac{x_{21-} x_{34-}}{x_{31-}x_{24-}} \;,\quad
v = -\frac{x_{14}^2 (n n')}{2x_{31-}x_{24-}} \;,\quad      \gamma =  \frac{  2 (nx_{14} ) (n' x_{14}) }{ x_{14}^2  \, (n n')} \,.
\end{align} 
The dimensionless and inert under the rescalings \p{rhon} variable $\g$ is the space-time counterpart of the angular variable $\zeta$ in \p{zetadef}. 
The function
$  \mathcal{G}(u,v , \g ) $ is given by 
 \begin{align}   
 \label{eq:3.5}
& \mathcal{G}(u,v , \g ) = \cP (u,v, \g) \,  \Phi^{(1)} (u,v)   + \frac{v^2}{u^3}  r_r ( u,v , \g)  \,,
\end{align} 
where  $ \cP$ is a  second-order differential operator, 
\begin{align}\label{3.8}
\cP ( u, v, \g) \equiv \frac{v^4}{u^2}  \left[  r_{uu} (u,v , \g) \,  \pa_{uu}^2 +  r_{uv} (u,v , \g) \,   \pa_{uv}^2 +  r_{vv} (u,v , \g) \,  \pa_{vv}^2  \right] 
\end{align} 
and $r_{uu}, r_{uv}, r_{vv} $ and $r_r$  are polynomials in $u,v, \gamma$:
\begin{align}\label{}
&r_{uu}= -u \left(2 u^3+u^2 (3-2 \gamma  v)+u ((4-3 \gamma ) v+3)+(\gamma -2) v^2-2 (\gamma -1) v+3\right) \,,  \nt
&r_{uv}=-2 u^4+u^3 (2 (\gamma +1) v+1)+u^2 v (\gamma -2 \gamma  v+3)-u v (\gamma +2 (\gamma -1) v-5)\nt
&\hskip1cm  +(\gamma -2) v^3+(4-3 \gamma ) v^2-2 \gamma  v+v+1  \,,   \\
&r_{vv}= v \left(-2 \gamma +2 u^3+u^2 (1-2 \gamma  (v+1))-(\gamma -3) u v+u+(\gamma -3) v^2+v+1\right)  \,,   \nt
& r_r = -u^3-u^2 (\gamma  v+v-2)+u \left((3 \gamma +4) v^2-(\gamma +1) v-1\right)+v \left((\gamma -2) v^2+2 v+1\right) . \nn
\end{align}
We see that the detector limit drastically simplifies the correlation function.

\subsection{Detector time integration}

At the next step, we need to analytically continue the Euclidean result \p{3.5} to Minkowski space with Wightman prescription    and then  integrate over the detector times. 
This defines the integrated charge-flow correlation  in position space, 
\begin{align}
\vev{J^\mu  \mathcal{Q} \mathcal{Q}  J_\mu }  &   =  -\frac{g^2 N_c C_F}{2^4 \pi^{10}}\frac{ (n  \bar{n} ) ( n'   \bar{n}')  }{x_{14}^8  \, (n  n')^3 }   
\int_{-\infty}^\infty  d  x_{2-} \, d  x_{3-}    \, \mathcal{G}(u, v, \g )\, .
\end{align}
We present two different approaches to computing the integrated correlator. In subsection~\ref{se321} we  adopt the Mellin approach \cite{Belitsky:2013xxa,Belitsky:2013bja}.  In subsection~\ref{se322} we explain and apply the method of  integrated double discontinuity \cite{Henn:2019gkr}.

\subsubsection{Analytic continuation via Mellin transform} \label{se321}

In Minkowski space, the space-time intervals $x_{ij}^2$  have small imaginary parts which reflect the Wightman prescription \p{1.19}, i.e. 
  $x_{21-} \to x_{21-} + i \ep$, $x_{24-} \to x_{24-} - i \ep$, $x_{31-} \to x_{31-} + i \ep$, $x_{34-} \to x_{34-}-i \ep$ .  
The  prescription specifies the integration contours over the detector time variables. 
The integration of the rational term in \eqref{eq:3.5}  is straightforward.  The contour integrals can be done by taking the residues at $x_{21-} = 0$ and $x_{34-} =0$. 

The first term in \eqref{eq:3.5} involves the second derivatives of $\Phi^{(1)}$.
It contains polylogarithms, whose analytic continuation can be tricky and the time integration appears to be more difficult.    
Following \cite{Belitsky:2013xxa,Belitsky:2013bja}, we first convert the   correlation function to Mellin space. The 
 analytic continuation of the Mellin transform  is straightforward \cite{Mack:2009mi} because  its space-time dependence is very simple. 
More specifically, we use the Mellin representation \cite{Usyukina:1993ch,Usyukina:1992wz}
\begin{align}
\Phi^{(1)}(u,v) = \int_{\delta -i\infty }^{\delta+i\infty} \frac{d j_1 d j_2 }{(2\pi i)^2} M(j_1,j_2) u^{j_1} v^{j_2}\,,
\end{align}
where $-1/2<\delta<0$ and  
\begin{align}\label{}
M(j_1,j_2) = - \frac14  [\Gamma(-j_1) \Gamma(-j_2)\Gamma(j_1 + j_2 +1)]^2\,.
\end{align}
The derivatives of $\Phi^{(1)}$ in \p{3.8} are equivalent to the following substitutions in the Mellin representation
\begin{align}
\pa^2_{uu} \to j_1(j_1-1) u^{-2} \;,\; 
\pa^2_{vv} \to j_2(j_2-1) v^{-2} \;,\; 
\pa^2_{uv} \to j_1 j_2 \, u^{-1} v^{-1} \,.
\end{align}
Schematically, we can write $\cP(u,v, \g) \Phi^{(1)}$  in terms of the following  Mellin integral, 
\begin{align}\label{3.13}
  [ \cP \Phi^{(1)} ] (u,v, \g ) & = \int_{\delta -i\infty }^{\delta+i\infty}  \frac{d j_1 d j_2 }{(2\pi i)^2} M(j_1,j_2) \, \left[  \cP(u, v,  \g)  u^{j_1} v^{j_2} \right]  \nn\\
  & = \sum_{m,n}  \int_{\delta -i\infty }^{\delta+i\infty} \frac{d j_1 d j_2 }{(2\pi i)^2} f_{m,n } (j_1,j_2, \g ) \, u^{j_1-m} v^{j_2-n} \,,
\end{align}
 where $m,n$ are integers whose range is determined by  the polynomials $r_{uv}, r_{uu},$ and $r_{vv}$. The crucial point is that the cross-ratios $u,v$ in \p{3.13} are immediately converted to the Wightman prescription above,  
 \begin{align}
u = \frac{ (x_{21-}  + i \ep ) (x_{34-} - i \ep)}{(x_{31-}+i \ep)(x_{24-} - i \ep)} \;,\quad
v = -\frac{ x_{14}^2 (n n')}{2 (x_{31-} - i \ep) (x_{24-} + i \ep)} \,.
\end{align}

The next step is to carry out the  time integration   with the help of the formula
\begin{align}\notag
& \frac{(n\bar n)}{2\pi i}  \, \int d x_{2-} (-x_{21-} - i \ep)^{p-1} (x_{24-} - i \ep)^{q-1} =  \frac{\Gamma(1-p-q)}{\Gamma(1-p)\Gamma(1-q)} ((n x_{14})-i\ep)^{p+q-1}\,,  
\end{align} 
and similarly for the detector point 3. 
We  obtain the Mellin representation for the integrated correlator
\begin{align}
\label{eq:3.14}
 &  \frac{ 2 (n \bar n) (n' \bar n' ) }{x^2_{14} (n n')}     \int_{-\infty}^\infty   \hskip-0.1cm  d x_{2-} d x_{3-}     [ \cP \Phi^{(1)} ] \nt  
  & = -4\pi^2\sum_{m,n}  \int_{\delta -i\infty }^{\delta+i\infty} \frac{d j_1 d j_2 }{(2\pi i)^2} f_{m,n } (j_1,j_2, \g ) \,  \left[ \frac{\Gamma(j_2-n -1 )}{\Gamma( m- j_1)\Gamma( j_1 + j_2-m-n )}  \right]^2 
    \gamma^{-j_2 +n+1 } \,.
    \end{align} 
 Summing up all the pieces, the integrand in Mellin space is $\g^{-j_2-1}$ times a polynomial in $\gamma$ of degree 4 with rational coefficients in $j_1,j_2$.
Implementing the Mellin integrations over $j_1, j_2$  and combining with the contribution from 
 the rational term $r_r(u, v , \g)$ in \p{eq:3.5}, we obtain the charge-charge correlation in coordinate space, 
\begin{align}
\vev{ J^\mu(x_1) \mathcal{Q}(n) \mathcal{Q}(n') J_\mu(x_4)} = \frac{N_c C_F g^2}{2\pi^{8} x_{14}^6\gamma^3 (nn')^2}\left[ {\rm Li}_2 \left( 1-\frac{1}{\gamma} \right) + \frac{3}{2}\log(\gamma) - \frac{\pi^2}{6} + \frac{\gamma}{2}  + \frac{7}{4}
\right]. \label{2.7.10}
\end{align}
It is a function of the variable $\g$ defined in \p{3.6} encoding the angular separation of the two detectors. The prefactor on the right-hand side gives it the required dimension and scaling weight under \p{rhon}.

\subsubsection{The method of double discontinuity} \label{se322}

The Mellin approach described in the previous subsection is based on the  Mellin representation of the correlation function.
At higher loop orders it involves the evaluation of multi-fold Mellin integrals (e.g., four-fold at two loops),
 thus making the calculation inefficient.   
In Ref.~\cite{Henn:2019gkr}, a different approach to the time integration was applied to the NNLO calculation of the energy-energy correlation in $\cN=4$ sYM where it proved to be very efficient.
 The idea is to convert the time integrals to the integrated  {\it double discontinuity} \cite{Caron-Huot:2017vep,Alday:2017vkk} of the correlation function. It allows us to perform both the analytic continuation and the  time integrals as one operation in position space, without referring to the explicit Mellin amplitude.  
Here we   generalize the method to the QCD correlation function and  provide some details not presented  in  \cite{Henn:2019gkr}.
 In particular, we explain how to derive the two-fold integral representation of $\vev{J  \mathcal{Q} \mathcal{Q}  J }$  in terms of the double discontinuity of $\vev{JJJJ}$. 
 In App.~\ref{dDisc} we list  the explicit rules for computing the double discontinuities of typical functions containing simple or higher-order poles.

Double discontinuity  is an operation that takes  twice the discontinuity of a multivalued function $g(w)$
  across  two adjacent Riemann sheets, 
\begin{align} 
& \text{dDisc}_{ w =0 }  \equiv  2\,  \text{disc}_{ w=0 }^{\circlearrowleft} \text{disc}_{ w=0 }^{\circlearrowright}\,,   \\
& \text{where} \hskip0.45cm 
\text{disc}^{\circlearrowleft}_{w=0}\,  g   \equiv   \frac{1}{2i } \left[ g (e^{2\pi i } w  )  - g  (w)  \right] ,  
\quad \text{disc}^{\circlearrowright}_{w=0}  \, g  \equiv   -  \frac{1}{2i } \left[ g (  e^{-2\pi i } w )- g (w) \right] \,.  \nonumber
\end{align} 
This definition implies  $\text{dDisc}_{w=0}[w^j ] = 2 \sin^2( \pi j)  \, w^j$, hence  the Mellin representations of a function  $g(w)$  and of  its double discontinuity 
 differ  by a factor of $2 \sin^2( \pi j) $. 
 
Now we show that the integrated correlation function can be defined through its double discontinuity. 
Let us assume that the weightless correlation function $\cG(u, v, \g)$ from  \eqref{3.5}  admits a Mellin representation,
\begin{align}
\label{eq:3.22}
 \cG  (u, v, \g )=  \sum_{m,n } \int \frac{d j_1}{ 2 \pi i} \frac{d j_2}{ 2 \pi i }  \, f_{m, n }  (j_1, j_2, \g)  \,  u^{j_1-m} v^{j_2-n} \,, 
\end{align} 
where the sum goes over integers within a certain finite range.  
The  two-fold time integral in \eqref{eq:3.14} evaluates to 
\begin{align} 
\label{eq:3.23} 
&  G (\g)   \equiv  \frac{ 2 (n \bar n) (n' \bar n' ) }{x^2 (n n')} 
 \int d x_{2-} d x_{3-}  \; \cG(u, v,\g)    \\
  &=\frac{1}{2 \pi^2} \sum_{m,n }  \hskip-.1cm 
   \int \frac{d j_1 d j_2 }{ (2 \pi i)^2}   \, f_{m, n }  (j_1, j_2, \g) 
 [2 \sin^2 (\pi j_1) ] \left[B(j_1-m+1, j_2-n-1)   \right]^2   
  \hskip-.1cm   \gamma^{-j_2 +n+1 } \,,   \nn 
\end{align} 
where we encounter  the Euler beta function 
\begin{align} 
B(j_1,j_2)=\frac{\Gamma(j_1)\Gamma(j_2)}{ \Gamma (j_1 + j_2) } = \int_{0}^1 \, dt\,   t^{j_2-1} (1- t)^{j_1-1} \,.
\end{align} 
The integral representation on the right-hand side  allows us to convert  the Mellin contour integrals  in \eqref{eq:3.23} into a two-fold integration over a finite interval, 
\begin{align}
\label{eq:3.25}
G (\g)  &   = - \frac{1}{2 \pi^2 }\frac{1}{\g}  \int_{0}^1 d t\,  d \bar t   \;  \bom{g} \Big( u (t, \bar t, \g) , v (t, \bar t, \g)  , \g \Big)   ,  \nn \\
 & \text{where} \quad   u (t, \bar t, \g)  \equiv  (1- t) (1-\bar t) , \quad  v (t, \bar t, \g)  \equiv  \frac{ 1}{\g}   t \bar t \,. 
\end{align}  
The function   $\bom g (u , v , \g)$ is defined through its Mellin representation, 
\begin{align}
\label{eq:3.26}
 \bom{g} (u , v , \g)    =   
  \sum_{m,n }\int \frac{d j_1 d j_2 }{(2\pi i)^2 }  \, f_{m, n }  (j_1, j_2, \g)  [ 2 \sin^2 \pi j_1  ] 
\,  u^{j_1 - m }  v^{j_2- n -2 } \,. 
\end{align} 
Comparing \eqref{eq:3.26} with \eqref{eq:3.22},   
 we observe that  $\bom g(u, v , \g) $ can be obtained by taking the double discontinuity of $\cG(u,v,  \g)$ at $u=0$,
\begin{align}
 \bom{g} (u  , v , \g)  = {\rm dDisc}_{u=0} \left[  \frac{1}{v^2}  \cG(u, v, \g)  \right] \,. 
\end{align} 
 
 The main advantage of this method is that, given a Euclidean correlation function, 
the double discontinuity can be worked directly without referring to its explicit Mellin representation.  
To proceed, we  set $u = (1-z) (1- \bar z)$, $v= z \bar z$, and 
change variables
\begin{align}\label{}
(z,\bz) \ \rightarrow \ ( t, \bar z) \quad \text{with} \ \ z  = \frac{t (t-\bar{z})}{\g t \bar{z}-\g \bar{z}-t \bar{z}+t}\,. 
\end{align} 
 Under such a reparametrization the integrand factorizes    linearly. 
We arrive at a new representation of  \eqref{eq:3.25}
 that expresses  the integrated correlator in terms of the double discontinuity at $\bar z =1$,
\begin{align}
 G(\g) & =    \int_{0}^1 d \bar z  \int_{0}^{\bar z} d t \,  \frac{1}{ \g \bar{z} (1-t) - t(1- \bar z )  }    \text{dDisc}_{\bar z =1 } \left[ \frac{(\bar z - z)  }{ (z  \bar z )^2 } \mathcal{G}(z, \bar z , \g) \right]  \,.
\end{align} 
The correlation function \p{Phi1} contains polylogarithms with branch cuts at $z, \bar z = 0 \; \text{or } 1 $. 
In order to take the double discontinuity, 
we extract the logarithmic singularities at $\bar z = 1$, so that the correlation function takes the form  
 \begin{align} 
 \frac{(\bar z - z)  }{ (z  \bar z )^2 } \mathcal{G}(z, \bar z , \g)=  a (z , \bar z, \g)  +  b (z, \bar z, \g)   \ln ( 1- \bar z )\,,    
 \end{align} 
 where  $a$ and $b $ are free from logarithmic singularities but they contain poles at $\bar z =1$ up to degree 3.
  We first carry out the integral  over $t$, which does  not increase the degree of the poles at $\bar z =1$ but produces additional powers of $\ln (1-\bar z)$. After this we are left with a one-dimensional integral, 
\begin{align} 
\label{eq:3.33}
G (\g) & =  \int_{0}^1 d \bar{z}\,    \bigg\{   {\rm dDisc}_{\bar z =1} \left[  B_0(\bar z , \g)  \ln (1- \bar z ) + A_0 (\bar z , \g) \right] \nn  \\
&  + \ln (1- \bar z )  \, 
 {\rm dDisc}_{\bar z =1} \left[  B_1(\bar z , \g)  \ln (1- \bar z ) + A_1 (\bar z , \g) \right] \bigg\}\,,  
\end{align} 
where $A_{0,1}$ and $B_{0,1}$ contain only isolated poles  but no branch cuts at $\bar z =1$.  
Taking the double discontinuity generates distributional terms at $\bar z =1$ (see Appendix \ref{dDisc} for detail).
The integral in  \eqref{eq:3.33} picks the residues of $A_1$ and $B_0$ at $\bar z =1$.  The  result is 
\begin{align}
\label{eq:3.34}
\vev{J^\mu  \mathcal{Q} \mathcal{Q}  J_\mu }  &   =  -\frac{N_c  C_F g^2}{2^5 \pi^{10}}\frac{ 1 }{x_{14}^6  \, (n  n')^2 }  G (\g)   \nn \\   
& =  -\frac{N_c C_F  g^2}{2^4 \pi^{8}}\frac{ 1  }{x_{14}^6  \, (n  n')^2 }    \left[  {\rm Res}_{\bar z =1} B_{0} (\bar z, \g) -   {\rm Res}_{\bar z =1} A_{1} (\bar z, \g) \right]  \nn \\
& = \frac{N_c C_F g^2}{2\pi^8 x_{14}^6\gamma^3 (nn')^2} \left[ {\rm Li}_2 \left( 1-\frac{1}{\gamma} \right) + \frac{3}{2}\log(\gamma) - \frac{\pi^2}{6} + \frac{\gamma}{2}  + \frac{7}{4}
\right] \,,
\end{align}
which fully agrees with the answer for the Mellin integral given in  \eqref{2.7.10}.

\subsection{Fourier transform and final result}

The last step in obtaining the event shape function $F_{\rm{QQC}} (\zeta)$ is converting the result \p{2.7.10} for the integrated correlation function  to momentum space according to the definitions \p{124} and \p{zetadef}. In our one-loop calculation the normalization factor $\sigma_{\rm tot}$   is evaluated  at Born level, since its finite $O(g^2)$ correction contributes  only to the contact terms in $F(\zeta)$. With the help of \p{theta} we find
\begin{align}\label{e3.29}
 \sigma_0 = \int d^4 x \, e^{iqx} \vev{J_{\mu}(x) J^{\mu} (0)}_{\rm Born}=\frac{2 N_c}{\pi^4} \int \frac{d^4 x\, e^{i q x}}{(-x^2 + i \epsilon x_0)^3} =\frac{N_c}{2\pi} q^2 \q(q^0) \q(q^2)\,.
\end{align} 
 The details of the Fourier transform of \eqref{2.7.10} are given in  Appendix \ref{FT}.   The result is 
\begin{align}
\label{final}
F^{\rm QCD}_{\rm{QQC}} (\zeta) & = 4\pi^2 (nn')^2 \sigma_{0}^{-1} \int d^4x\, e^{iqx} \,  \vev{ J^\mu(x) \mathcal{Q} \mathcal{Q} J_\mu(0)}  \nt
&= \frac{C_F g^2}{4\pi^2}   \frac{2\log(1-\zeta) +\zeta (2+\zeta)}{\zeta(1-\zeta)}  + O(g^4)\,.
\end{align}
This is the main result of the present paper. 
In Appendix~\ref{ampcalc} we obtain   the same result by the standard amplitude method.

 {For comparison we quote the $\mathcal{N} =4$ sYM result \cite{Belitsky:2013bja}
 \begin{align}\label{317}
F^{\cN=4}_{\rm{QQC}} (\zeta)
&  = \frac{g^2 N_c}{\pi^2} \, { \ln(1-\zeta)}
\left({2\zeta\over 1-\zeta} R_1+R_2 \right) + O(g^4)\,,
\end{align}
where  $R_1, R_2 $  are constant R-symmetry factors. We see that the leading asymptotic singularity when $\zeta \to 1$ (back-to-back scattering) in \p{final} and in \p{317} is the same, up to numerical coefficients which differ due to the R-symmetry factor.  The one- and two-loop results for the energy-energy correlations in $\cN=4$ sYM and QCD coincide in the back-to-back regime  \cite{Belitsky:2013bja,Belitsky:2013ofa}. }

In conclusion we recall that the complete result for the event shape includes contact terms $C_1 \delta(\xi)+ C_2 \delta(1-\xi)$ corresponding to the  collinear and back-to-back regimes. These contributions are beyond the scope of the present paper. For more detail see \cite{Belitsky:2013xxa,Kologlu:2019mfz,Korchemsky:2019nzm,Dixon:2019uzg}.

\section{Conclusion and future directions}\label{s5}

In this paper, we provided a proof of principle that event shapes in QCD can be
obtained from correlation functions. Contrary to the standard approach using
amplitude methods, where one encounters in general infrared divergences
in the virtual contributions and phase space integrals, our setup is 
completely infrared finite. This is a conceptual simplification.

In the paper we restricted ourselves to the  simplest case of the one-loop charge-charge correlation. The next logical step is to go to the physically more interesting case of the energy-energy correlation. It  requires the calculation of the four-point correlation function  of two currents  (source and sink) and two energy-momentum tensors (detectors of energy or calorimeters). This correlation function will have a more complicated tensor structure. Moreover, the energy-momentum tensor involves both the gauge field and the fermions, so we will have new types of  Feynman diagrams. We plan to develop the necessary tools for such calculations. 

At the order we computed, the traditional amplitude approach is without doubt more efficient.
However, this may very well change at higher orders in perturbation theory.\footnote{The QQC is known to be infrared divergent beyond one loop \cite{Kunszt:1992tn,Banfi:2010xy} but other observables, such as the EEC, are IR safe. } We may profit from the  well developed methods for computing correlation functions in
position space, see e.g. \cite{Drummond:2013nda} for three-loop results of a scalar four-point function in the $\cN=4$ sYM 
theory, certain higher-loop four-point integrals \cite{Eden:2016dir}, or results on four- and five-loop anomalous dimensions
from two-point functions \cite{Velizhanin:2014zla} and splitting functions \cite{Herzog:2018kwj}.
Furthermore, additional simplifications can be expected due to the finiteness
of our method, e.g. via special four-dimensional integral identities or methods \cite{Drummond:2006rz,Schnetz:2013hqa}.

We used conformal symmetry as an organizing principle for the result
of the four-point correlator we evaluated. This was very useful, but not essential
to our approach. In particular, the method of double discontinuity of Sect.~\ref{se322} is universal, it can be applied to a non-conformal correlation function. 
One important change at higher loops in QCD is the appearance of a non-vanishing  beta 
function and renormalization of the fields.\footnote{The conserved currents and energy-momentum tensor are protected from UV renormalization.} This means that UV divergences
will have to be dealt with. 
It also means further, non-conformal, structures can appear in the result for the 
correlator. It may be very interesting to study such terms at a conformal fixed point, see \cite{Braun:2003rp,Grozin:2015kna,Braun:2018mxm}
and references therein for related work.

An interesting question arises when comparing the correlator $ \vev{J\cQ \cQ  J  }$ to the full four-point function $\vev{JJJJ}$. In Sect.~\ref{s3.1} we have seen a dramatic simplification when two of the currents become charge detectors $\cQ$. What is the deep reason for this? The answer may come from comparing the OPE of currents to the OPE of light-ray operators (for a recent study see \cite{Kologlu:2019mfz}). It appears that the latter captures only a small subset  of the former. 

Another direction for further application of the method is the study of multi-detector energy correlations (for some very recent  results see \cite{Chen:2019bpb}).  In our language it will require computing correlation functions of the type $\vev{JT \ldots T J}$ with multiple energy-momentum tensor insertions. Already a Born level calculation can give us   non-trivial information about such highly complex observables.

\section*{Acknowledgments}

We are indebted to G. Korchemsky and A. Zhiboedov for numerous discussions. E.S. is grateful to the MPP-Munich for hospitality during the work on this project. This research received funding from the European Research Council (ERC) under the European Union’s Horizon 2020 research and innovation programme {\it{Novel structures in scattering amplitudes}} (grant agreement No 725110).

\appendix

\section{Conventions and conformal properties in position space}
\label{conv}

We use the two-component spinor conventions of \cite{Galperin:1984av,Galperin:2001uw}. The relations between Lorentz four-vectors and $2\times 2$ matrices are defined by
\begin{align}\label{}
x_{\a\da}=x^\mu(\sigma_\mu)_{\a\da}\,, \qquad \tilde x^{\da\a} = x^\mu(\tilde\sigma_\mu)^{\da\a} = \ep^{\a\b} \ep^{\da\db} x_{\b\db}\,,
\end{align}
with the sigma matrices $\sigma_\mu = (1,\vec\sigma)$ and $\tilde\sigma_\mu = (1,-\vec\sigma)$.
To raise and lower two-component indices we use  the Levi-Civita tensors
\begin{align}\label{}
\ep_{12}=-\ep^{12}=
\ep_{\dot{1}\dot{2}}=-\ep^{\dot{1}\dot{2}}=1\, ,\qquad \ep^{\a\b} \ep_{\b\gamma}=\delta^\a_\gamma  \,,
\end{align}
satisfying the  identities
\begin{align}\label{}
x_{\a\da} \tilde y^{\da\b} + y_{\a\da} \tx^{\da\b} = 2( x \cdot y) \delta_\a^\b\,, \qquad\;
x_{\a\da} \tx^{\da\b} = x^2 \delta_\a^\b\,, \qquad 
x^2 = \frac1{2}  x_{\a\da} \tx^{\da\a}\,.
\end{align}
The space-time derivative is defined as $\pa_{\a\da} = \sigma^\mu_{\a\da} \pa_\mu$ and has the property
\begin{align}\label{A4}
\pa_{\a\da} \tx^{\db\b} = 2 \delta_\a^\b \delta_\da^\db \,.  
\end{align}

The easiest way to check conformal invariance is to make the discrete operation of conformal inversion
\begin{align}\label{}
I[x^\mu] =\frac{x^\mu}{x^2}  \,, \qquad I^2 = \mathbb{I}\,.
\end{align}
The basic fields in a $D=4$ conformal theory transform with specific conformal weights:
\begin{align}\label{B2}
I[\phi] =   x^2 \phi\,, \quad I[\psi_\a] = x^2 \tx^{\da\a} \psi_\a\,, \quad I[\bar\psi^\da] = -x^2 x_{\a\da} \bar\psi^\da\,, \quad I[F_{\a\b}] =x^2 \tx^{\da\a} \tx^{\db\b} F_{\a\b} \,,
\end{align}
namely $(+1)$ for a scalar $\phi$, $(+3/2)$ for a spinor $\psi$ and $(+2)$ for a field strength. 
These weights are chosen so that the field equations are covariant. For example, let us check the covariance of the free Maxwell equation $\tilde\pa^{\da\a} F_{\a\b}=0$, where $F_{\a\b}= F_{\b\a}$ is the self-dual half of the Maxwell field strength. To this end we use the inversion property of the derivative 
\begin{align}\label{B3}
 \pa_{\a\da}  \ \stackrel{I}{\rightarrow} \  -\tx^{\da\b} \tx^{\db\a} \pa_{\b\db} \,,
\end{align}
as follows  from the identity
\begin{align}\label{}
 \pa_{\a\da} x^2 = 2 x_{\a\da}  \ \stackrel{I}{\rightarrow} \   -\tx^{\da\b} \tx^{\db\a} \pa_{\b\db} \left(\frac1{x^2}\right) = 2 \frac{x_{\a\da}}{x^2}\,.
\end{align}
We then have
\begin{align}\label{}
\tilde\pa^{\da\a} F_{\a\b} \ \stackrel{I}{\rightarrow} \  &  -x_{\a\dot\rho} x_{\rho\da} \tilde\pa^{\dot\rho\rho}(x^2 \tx^{\da\gamma} \tx^{\db\delta} F_{\gamma\delta} ) 
= -x^4 x_{\a\dot\rho}\tx^{\db\delta} (\tilde\pa^{\dot\rho\gamma} F_{\gamma\delta}) \,,
\end{align} 
where we have used the property $\ep^{\a\b} F_{\a\b}=0$ of the self-dual tensor.   Introducing a vector source term (conserved current) $j^\da_\b$ of  the Maxwell equation, we deduce its inversion law
\begin{align}\label{B.6}
\tilde\pa^{\da\a} F_{\a\b} = j^\da_\b \ \Longrightarrow \ j^\da_\b  \ \stackrel{I}{\rightarrow} \    -x^4 x_{\a\dot\rho} \tx^{\db\delta}\, j^{\dot\rho}_{\delta}\,.
\end{align} 
The current transforms as a vector with conformal weight $(+3)$ which is the sum of the weights of the derivative and the field strength.

\section{Charge-charge event shape  from LO amplitude calculation}\label{ampcalc}

As explained in Sect.~\ref{s1}, the standard approach to event shapes is to 
start from their definition  as  weighted cross sections.
 One computes the matrix element $ |\cM (X_i \rightarrow X_f)|^2$, sums over all the final states $f$ and integrates over the phase space,  
\begin{align} 
\label{Eshape}
 F_w (\zeta) \equiv \sum_{X_f}  \hskip-0.1cm \int  \hskip-0.1cm {\rm d} \Pi_f  \, |\cM (X_i \rightarrow X_f)|^2 \,  {w}(\zeta;  X_f) \,,
\end{align} 
 where ${w}$ is a measurement function that determines how to weigh a given  final state.  In the case of two detectors  the event shape is a function of the variable $\zeta$ defined in \p{zetadef} and related to the angle $\q$ between the two detectors. 
The calculation can be done in the center-of-mass frame $q^\mu=(q^0,\vec 0)$, where the measurement function for charge correlations is given by 
 \begin{align} 
 \label{wQQC}
 w_{\rm QQC} ( \zeta; X_f) & \equiv  16 \pi^2 \zeta^2  \sigma_{\rm tot}^{-1} \hskip-0.1cm \sum_{i, j } \hskip-.05cm  Q_i Q_j  \int d^2 \vec{n} \, d^2 \vec{n}' \, \delta \big(  2 \zeta -  (n n')  \big) \nn \\
& \hskip+4cm 
 \times  \delta^2 (\Omega_{\vec{k}_i} - \Omega_{\vec{n}}) \, 
 \delta^2 (\Omega_{\vec{k}_j} - \Omega_{\vec{n}'})\,.
\end{align}
Here $ n = (1, \vec{n}), n'=(1, \vec{n}')$ and  $\vec{n}, \vec{n}'$ are the unit vectors defining the directions of the detectors.   The overall factor in \p{wQQC} comes from our definition \p{zetadef} of the event shape function $F(\zeta)$ (recall   \p{1.27}). 
In \p{wQQC}, we sum over $i$ and $j$ running over all particles in the final state,
and we integrate over the orientations of the detectors, fixing the angle between them.  
The charge correlation defined through \p{Eshape} and \p{wQQC} is consistent with the standard definition of 
a similar event shape, namely EEC 
(see e.g. \cite{Ellis:1991qj,Ellis:1980wv}).
More explicitly, 
\begin{align} \label{B.3}
F_{\rm QQC}(\zeta)  \equiv 
   2 \zeta^2  \sum_{i, j}  \int  d \cos \theta_{ij} \, Q_i Q_j  \,  \delta \Big(  2 \zeta - 1 + \cos \theta_{ij}   \Big)  \,   \frac{1}{\sigma_0} \frac{ d \sigma}{ d \cos \theta_{ij}}\,.
\end{align}
The definition here is slightly different from \p{wcs}  where $\vec{n}, \vec{n}'$ are fixed without being integrated over. 
This results in an extra overall factor,  
  $  \int d^2 \vec{n} \, d^2 \vec{n}' \, \delta \big(  2  \zeta -  (n n')  \big)  \,  F(\zeta)  = 8 \pi^2 \,  F (\zeta) $, which has been incorporated in the normalization in  \p{B.3}.

\begin{figure}[ht!]
    \centering
    \begin{subfigure}[c]{0.2\textwidth}
      \includegraphics[scale=0.4]{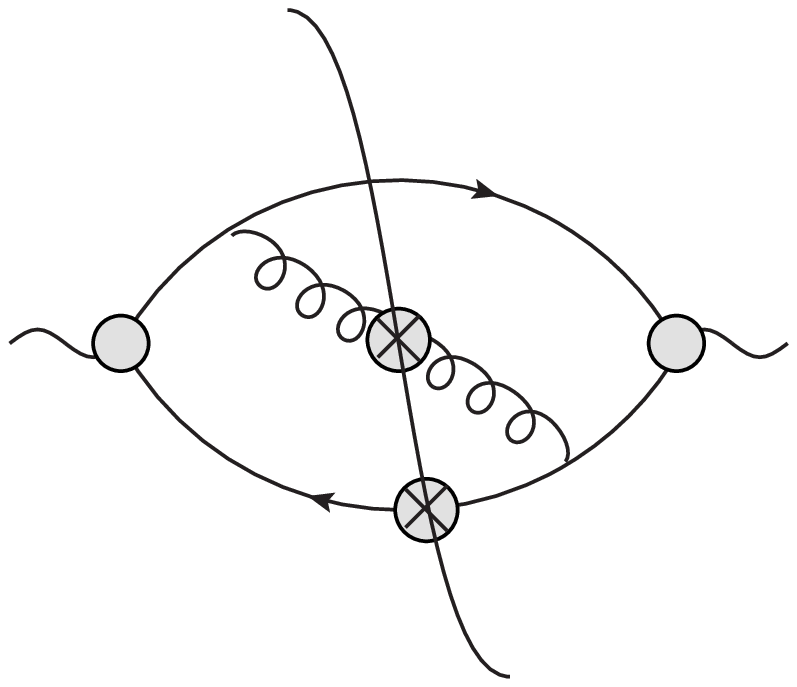}
        \caption{}
        \label{fig:rb}
    \end{subfigure}
  \quad
    \begin{subfigure}[c]{0.2\textwidth}
      \includegraphics[scale=0.4]{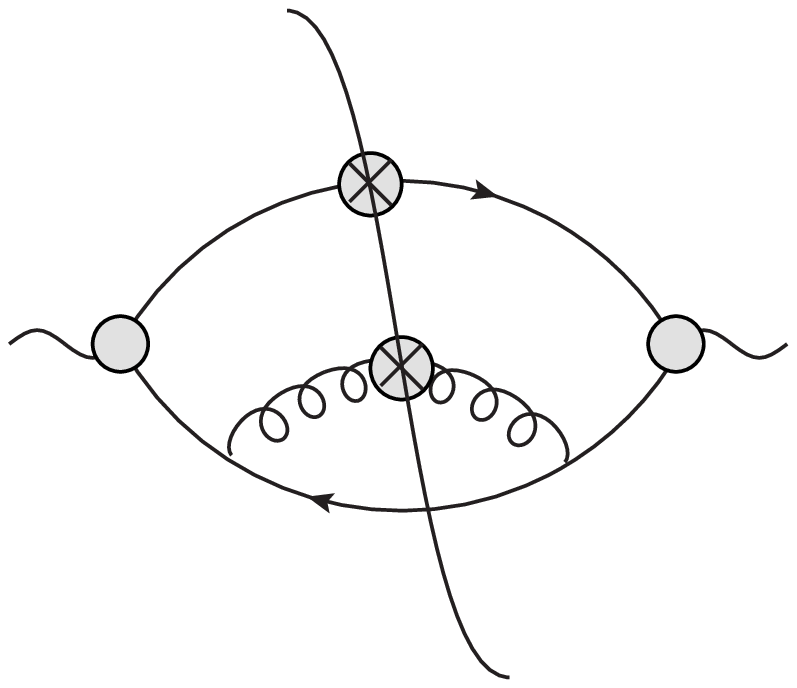}
        \caption{}
        \label{fig:rc}
        \end{subfigure}
      \quad 
    \begin{subfigure}[c]{0.2\textwidth}
      \raisebox{0pt}{\includegraphics[scale=0.4]
         {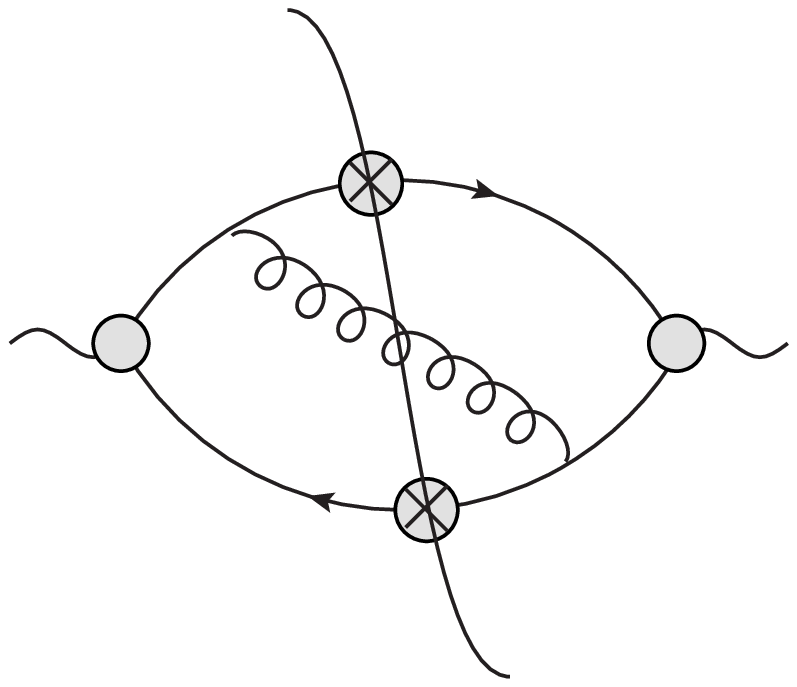}}
        \caption{}
        \label{fig:rd}
    \end{subfigure}    
    \quad
    \begin{subfigure}[c]{0.2\textwidth}
      \includegraphics[scale=0.4]{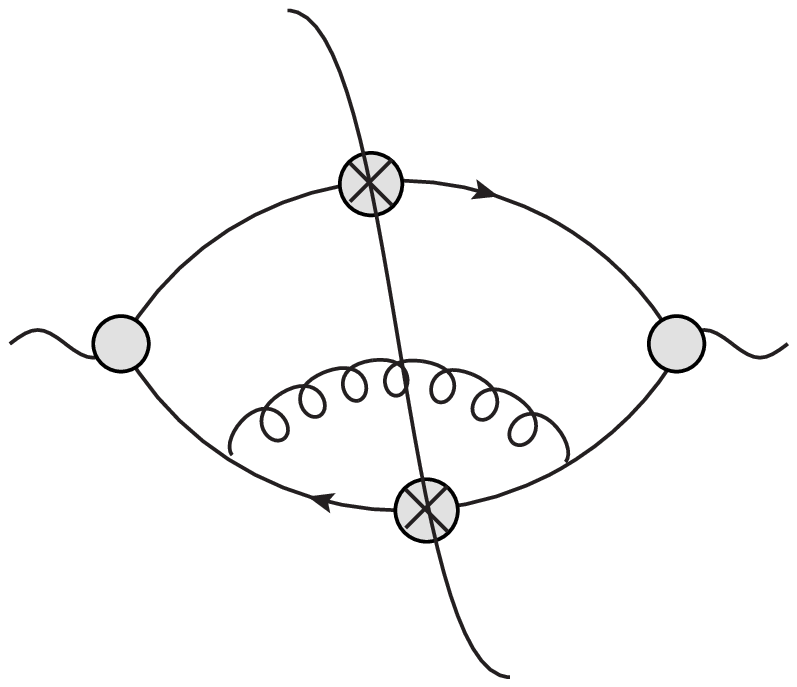}
        \caption{}
        \label{fig:re}
    \end{subfigure}
    \caption{ 
  Representative  cut real emission Feynman diagrams for the  one-loop  cross section  in $e^{+}e^{-}$  annihilation where the final-state particles penetrate the two detectors. In (a) and (b)  the detectors capture the gluon and (anti-)quark; 
      in (c) and (d)  they capture the  quark and anti-quark.}   
\label{fig:region} 
\end{figure}
 We are interested in the charge-charge correlation at one-loop order,  where the detectors are separated by a generic angle. 
 In this case we do not consider the situation where both detectors are aligned and capture the same particle, i.e. if $\zeta =0$. 
At $\zeta \neq 0$, we also do not consider the pure virtual diagrams at this order. 
Their contribution is a contact term  $\sim \delta(\zeta -1)$,  
 since momentum conservation forces the two final-state particles to go back to back. 
For this reason we will focus on the process  $\gamma^*(q) \rightarrow   {q} (k_1)   {\bar q} (k_2 )  {g} (k_3) $ at tree level, whose contribution is a regular function for $0< \zeta<1$. First,  we compute the matrix element of the initial state sourced by the vector electromagnetic current   $J_ \mu =  \bar{\Psi} \gamma_{\mu } \Psi  $ (with $\Psi, \bar\Psi$ standing for the quark/antiquark fields) and the three-particle final state.
After summing over the final state helicities and taking the color trace,  the matrix element squared reads 
\begin{align} 
\label{eq:4.1}
&  \sum_{\rm{helicity}} {\rm tr} \,
 |\cM_\mu|^2  = { \sum_{\rm{helicity }}  \rm tr} \,  \big|  \bra{0} J_\mu (0)    \ket{ k_1\,  k_2 \, k_3 }  \big|^2   \nn \\
& \hskip.5cm  =   { 32\pi }   { \alpha_s} N_c C_F  \, Q_f^2  \left[ \frac{s_{23} }{s_{13} } + \frac{s_{13}}{s_{23 }} +  \frac{ 2 q^2 s_{12} }{ s_{13} \,  s_{23}  }  \right] \,, 
\end{align} 
where  $\alpha_s ={g^2}/({4 \pi})$, and $Q_f$ is the (fractional) charge carried by a quark of a given flavor $f$. 
Let us introduce the dimensionless  variables $\tau_i ,  \; i = 1, 2,3 $,
\begin{align}
& \hskip.5cm   \tau_i   \equiv \frac{   2 q \cdot k_i  }{  q^2 } , \quad     
   \text{so that} \quad  \frac{  s_{ij}}{q^2}   =   1- \tau_n  ,     \;\;     n \neq i \neq j,  \quad     \text{and} \quad      \tau_1 + \tau_2 + \tau_3 = 2  \, . 
\end{align} 
In the center-of-mass frame  
 $\tau_i$ defines the fraction of the total energy carried by the $i$th particle.  
The three  $\tau-$variables also determine the  angles between each pair of final-state particles. More explicitly,  
\begin{align}  
\label{eq:4.4}
 \frac{2 E_i}{q^0} = \tau_i \, , \quad  \; \frac{1- \cos \theta_{ij} }{2}   = \frac{ 1- \tau_n }{\tau_i \tau_j} , \quad  n \neq i \neq j  \,. 
\end{align}
The three-body final-state phase space is conveniently parametrized by the $\tau_i$'s,
\begin{align} 
\label{eq:4.6}
 & {\rm d } \Pi_{3} \equiv    (2 \pi)^4 \delta^4(q- k_1 -k_2 -k_3)  \prod_{i=1}^3 \frac{d^4 k_i}{ ( 2 \pi)^4}   2\pi \delta_{+} (k_i^2)   \nn \\
 & = \frac{q^2}{ (4\pi)^5} 
      d \Omega_1 d \phi_{12}
       \, d \tau_1 d \tau_2  d \tau_3 \,   \delta (2 - \tau_1- \tau_2 - \tau_3 )    \prod_{i=1}^3  \theta (\tau_i) \theta(1- \tau_i) \,,
\end{align}
where $\Omega_1$ is the direction of $\vec{k}_1$, and $\phi_{12}$ is the azimuthal angle of $\vec{k}_2$ in the coordinate system where  $\vec{k}_1$ is the polar axis. To proceed, we combine the matrix element  \eqref{eq:4.1} and  the phase-space measure \eqref{eq:4.6},  and thus define the  tree-level differential cross section for  $\gamma^* \rightarrow    {q}_f     {\bar q}_f   {g} $ ($q_f/\bar{q}_f$ stands for (anti-)quark of a given flavor),
\begin{align} 
\frac{1}{\sigma_0}  \frac{d \sigma^f_{\rm LO}}{ d \tau_1 d \tau_2 } =   \frac{\alpha_s}{ 2\pi} \, C_F \, \frac{ Q_f^2}{\sum_f Q_f^2} \frac{\tau_1^2 + \tau_2^2  }{ (1- \tau_1) (1- \tau_2)  } \,.
\end{align}
Here   $\sigma_0 =  \frac{  N_c}{ 2 \pi } \sum_f  Q_f^2 \, q^2 $  
  is the Born-level total cross section where we sum over the flavors of the quarks. 

Now we are ready to compute the charge-charge correlation. 
By definition,  it measures the differential cross section weighted by the electric charge carried by particles $ i$ and $j$ at a fixed angular separation (recall \p{B.3}),   
\begin{align} 
F_{\rm{QQC}}( \zeta )  
& \stackrel{\rm{LO}}{ =}       2 \zeta^2  \sum_f
\int_0^1 d \tau_1 \int_{1-\tau_1}^{1} 
\hskip-0.2cm 
d \tau_2 \,  \delta \Big( \zeta  -  \frac{ \tau_1 +\tau_2 -1 }{ \tau_1 \tau_2 } \Big)  
\cdot  (-Q_f) \cdot (Q_f) \cdot  \frac{1}{\sigma_0} \frac{d \sigma_{\rm LO}^f }{ d \tau_1 d \tau_2 }  \,. \notag
\end{align}
Here we sum over  unordered pairs $(ij)$ in the final state. 
The two-fold phase-space integral   yields the one-loop charge-charge correlation 
\begin{align} \label{C8}
 F_{\rm{QQC}}( \zeta )   & \stackrel{\rm{LO}}{ =}  \frac{\alpha_s}{\pi}  \, C_F \,  R_Q  \frac{2 \log (1-\zeta  )  + \zeta  (2+\zeta )}{ \zeta (1- \zeta )}\,,
\end{align} 
where $R_Q \equiv \frac{\sum_f Q_f^4}{ \sum_{f} Q_f^2}$. This result agrees with \eqref{final},\footnote{In the correlation function approach the charge detector is made from the conserved electromagnetic current $J_\mu$, recall \p{Q-flow}. In this definition we have not specified the units of charge for a given particle, hence the absence of the factor $R_Q$ in \eqref{final}.}  
 thus showing the consistency between the amplitude calculation and our approach based on  correlation functions.

\section{Tensor reduction of the one-loop integrals}
\label{tens-red}

In order to evaluate the integrals with tensor numerators we need to reduce them to scalar integrals. We apply the following procedure for the tensor reduction.  The most complicated tensors that we encounter have rank two. Poincar\'e invariance implies that the tensor integral is a sum of 10 tensor structures,\footnote{Given 4 points $x^\mu_i$,  there are 3 independent translation invariant variables $x^\mu_{ij}$, from which we can make $3^2=9$ rank-two tensors. The 10th tensor is the Kronecker $\delta^{\mu \nu}$.}
\begin{align}
\int d^4 x_0 \frac{x^\mu_{0 a_1}\,x^\nu_{0 a_2} }{(x^2_{01})^{n_1}\,(x^2_{02})^{n_2}\,(x^2_{03})^{n_3}\,(x^2_{04})^{n_4}} = 
\sum_{k \neq a_1, \,l \neq a_2} x_{k\,a_1}^\mu x_{l\,a_2}^\nu \, I_{k,l} + \delta^{\mu \nu} \, I_0 \label{rank2}\,,
\end{align}
where $I_{k,l}$ and $I_0$ are some scalar integrals.
To find them we define the projectors
\begin{align}
P^\mu_{k;a} =& x_{k a}^\mu\, [  (x_{i_1 a} x_{i_2 a})^2 - x_{i_1 a}^2 x_{i_2 a}^2 ] + x_{i_1 a}^\mu \, [  (x_{k a} x_{i_1 a}) x_{i_2 a}^2 - (x_{i_1 a} x_{i_2 a}) (x_{k a} x_{i_2 a}) ] \notag\\
&+ x_{i_2 a}^\mu \, [  (x_{k a} x_{i_2 a}) x_{i_1 a}^2 - (x_{i_1 a} x_{i_2 a}) (x_{k a} x_{i_1 a}) ]\,,
\end{align}  
where $k \neq a$  and $\{i_1,i_2,k,a\}$ is a permutation of $\{ 1,2,3,4 \}$.
They satisfy 
\begin{align}
P_{k;a} \cdot x_{i a} = 0  \qquad \text{for} \;\; i \neq k\,.
\end{align}
Then we define the set of 10 rank-two projectors  
\begin{align}\label{C4}
& T^{\mu \nu}_{k,l} = 
P_{k;a_1}^\mu \, P_{l;a_2}^\nu \;,\qquad k \neq a_1\;,\quad l \neq a_2\,; \notag\\
& T^{\mu \nu}_{10} =
-\delta^{\mu\nu}  + \sum_{k \neq a_1, \, l \neq a_2}  \frac{(x_{k a_1} \cdot x_{l a_2})}{(P_{k;a_1}\cdot x_{k a_1})(P_{l;a_2}\cdot x_{l a_2})} P_{k;a_1}^\mu P_{l;a_2}^\nu \,.
\end{align}
The projector $T_{10}$   has the useful properties
\begin{align}
T_{10}^{\mu\nu} \delta_{\mu\nu} = 3-D \;,\qquad T_{10}^{\mu\nu} \, (x_{k a_1})_{\mu} (x_{l a_2})_{\nu} = 0 \,.
\end{align}

Now we project eq.~\p{rank2} with the 10 projectors,
\begin{align}
& \int d^4 x_0  \frac{x^\mu_{0 a_1}\, (T_{k, l})_{\mu\nu}\,x^\nu_{0 a_2} }{(x^2_{01})^{n_1}\,(x^2_{02})^{n_2}\,(x^2_{03})^{n_3}\,(x^2_{04})^{n_4}} = 
x_{k\,a_1}^\mu (T_{k,l})_{\mu \nu}\,  x_{l\,a_2}^\nu \, I_{k,l} + T_{k,l}^{\mu \nu} \delta_{\mu \nu} \, I_0 \;,\notag \\
& \int d^4 x_0  \frac{x^\mu_{0 a_1}\, (T_{10})_{\mu\nu}\,x^\nu_{0 a_2} }{(x^2_{01})^{n_1}\,(x^2_{02})^{n_2}\,(x^2_{03})^{n_3}\,(x^2_{04})^{n_4}} = 
(3-D) \, I_0 \,.
\end{align}
This triangular system of equations is solved immediately and we find the scalar integrals $I_{k,l}$ and $I_{10}$.

\section{Rules for taking  double discontinuity}
\label{dDisc}

Starting from the master formula,  $\text{dDisc} [w^{-p+\ep}] = 2 \sin^2 (\pi \ep ) w^{-p+\ep}  $  
, we can generate  logarithms terms via derivatives and work out the double discontinuity 
of pole times powers of logarithms (e.g. the integrand in \p{eq:3.33})  efficiently.
For $p, n , m  \in N$,   the general formula is
\begin{align}
\label{D1}
& \text{dDisc}_{w=0} [ w_+^{-p+\ep} \ln^m w_+  ] \ln^n w_+   \nn \\
 & =
\partial^m_\ep \left[  2 \sin^2 (\pi \ep )\,\partial^n_\ep \,  \frac{\Gamma( \ep-p+1 ) }{\Gamma( \ep  )} \,  \partial_{w}^{p-1} \hskip-0.1cm   \left(\frac{1}{\ep } \delta(w ) +  \sum_{k=0} \frac{\ep^k}{k!} w^{-1}_+ \ln^k w_+ \right)  \right] \,.  
\end{align}
Taking the limit $\ep\to0$, we obtain the double discontinuity of a $p-$th degree pole times multiple powers of logarithms. 
Here are a few examples of the application of \p{D1} for $n=0,1$, $m=0,1$:
\begin{align} 
\label{D2}
&\text{dDisc}\, w_+^{-p+\ep}    \ \stackrel{\ep\to0}{\longrightarrow} \ 0\,,\nt
& \text{dDisc} [ w_+^{-p+\ep}] \ln w_+     \ \stackrel{\ep\to0}{\longrightarrow} \  -2\pi^2 \frac{(-1)^{p-1}}{(p-1)!} \, \delta^{(p-1)}(w)\,,\nt
& \text{dDisc} [ w_+^{-p+\ep} \ln w_+  ]      \ \stackrel{\ep\to0}{\longrightarrow} \  2\pi^2 \frac{(-1)^{p-1}}{(p-1)!}  \, \delta(w)\,, \nt
& \text{dDisc} [ w_+^{-p+\ep} \ln w_+  ] \ln w_+  
\ \stackrel{\ep\to0}{\longrightarrow} \    0\,.   
\end{align} 
Applying these formulas to the integral in \eqref{eq:3.33}  we find
\begin{align} 
G (\g) & =   2\pi^2 \sum_{p}  \frac{1}{ (p-1)! } \left[ \pa_{\bar z }^{p-1}  B_0 (\bar z , \g) -  \pa_{\bar z}^{p-1} A_1 (\bar z , \g)  \right] \bigg|_{\bar z =1}   \nn \\
 & = 2\pi^2  \,  \text{Res}_{\bar z =1}  \left[ B_0 (\bar z , \g)- A_1 (\bar z , \g)  \right] \,,
\end{align}  
as stated in \p{eq:3.34}.

\section{Fourier transform in eq.~\p{final}}\label{FT}

We start with the Fourier integral\footnote{In what follows we assume $q_0 >0$ and $q^2 >0$.} 
\begin{align}
\int d^4 x  \frac{e^{i q x}\gamma^{j}}{(x^2-i\ep x^0)^3 \gamma^3} = -\frac{\pi^3 q^2 \zeta ^{3-j}}{2\Gamma(j)\Gamma(5-j)}\,  {}_{2} F_1 \left( 2-j,2-j;5-j|\zeta  \right) \label{Fourier2}\,,
\end{align} 
with $\g$ and $\zeta$ defined in \p{3.6} and \p{zetadef}, respectively. This integral is 
obtained using Schwinger's parametrization, with the help of the  formula \cite{Belitsky:2013xxa}
\begin{align}
&\int_0^{\infty} d\omega d\omega'  \,(\omega\omega')^a \left[(q-n\omega - n'\omega')^2\right]^b \theta((q-n\omega - n'\omega')^2) \theta(q_0-n_0\omega - n'_0\omega') \notag\\ 
&= \frac{\Gamma^2(a+1)\Gamma(b+1)}{\Gamma(2a+b+3)} \zeta ^{a+1} (q^2)^{a+b+1} (2 (nn'))^{-a-1}
\,  {}_{2} F_1 \left( a+1,a+1;2a+ b + 3|\zeta  \right)\,.
\end{align}
Using \p{Fourier2} we find 
\begin{align}
& \int d^4 x  \frac{e^{i q x}}{(x^2-i\ep x^0)^3 \gamma^3} = 0\,, \nt
& \int d^4 x  \frac{e^{i q x} \gamma}{(x^2-i\ep x^0)^3 \gamma^3} = -\frac{\pi^3 q^2}{2 \zeta } [(\zeta -2)\log(1-\zeta )-2\zeta ]\,, \notag\\
& \int d^4 x  \frac{e^{i q x} \log(\gamma)}{(x^2-i\ep x^0)^3 \gamma^3} = -\frac{\pi^3 q^2}{8 \zeta (1-\zeta )} [ \zeta (6 - 5\zeta ) + (2\zeta ^2-8 \zeta  + 6)\log(1-\zeta )]\,.
\end{align} 
To obtain the last identity we acted with $\pa_j|_{j = 0} $ on both sides of \p{Fourier2}.

The Fourier transform of the dilogarithm term in eq.~\p{2.7.10} is found with the help of the relation \cite{Korchemsky:2015ssa}
\begin{align}
\left[\gamma^2(1-\gamma)^2 G^{''}(\gamma) \right]^{''} = \frac{1}{16} (x^2)^3 \Box^2_{x} \frac{G(\gamma)}{x^2}\,.
\end{align}
We substitute $G(\gamma) = \frac1{\g} {\rm Li}_2 \left(1-\frac{1}{\gamma}\right)  $ in this equation and Fourier transform both sides,
\begin{align}
&\int d^4 x  \frac{e^{i q x}}{(x^2-i\ep x^0)^3 \gamma^3} \left[ 4 {\rm Li}_2 \left(1-\frac{1}{\gamma}\right) +12 \log(\gamma) + 6 \gamma-13\right]\nt
& = \frac{(q^2)^2}{16} 
\int d^4 x  \frac{e^{i q x} }{x^2-i\ep x^0} \frac1{\g}  {\rm Li}_2 \left(1-\frac{1}{\gamma}\right) = \frac{\pi^3 q^2}{2} \frac{\zeta \log(1-\zeta )}{1-\zeta } \,,
\end{align}
where the second relation can be found in  \cite{Belitsky:2013xxa}. Thus we get
\begin{align}
\int d^4 x  \frac{e^{i q x}}{(x^2-i\ep x^0)^3 \gamma^3} {\rm Li}_2 \left(1-\frac{1}{\gamma}\right) = \frac{\pi^3 q^2}{2} \frac{3(2-\zeta )\zeta  + (6-6\zeta +\zeta ^2)\log(1-\zeta )}{\zeta (1-\zeta )} \,.
\end{align}

Combining these steps yields \p{final} in the main text.

\bibliographystyle{JHEP}

\end{document}